\documentclass[twocolumn,showpacs,preprintnumbers,amsmath,amssymb]{revtex4}
\usepackage{graphicx}
 \usepackage{psfrag}
 \usepackage{epsfig}
 
\begin{document}

 \title{Response and fluctuations of a two-state signaling module with
feedback}
 
 \author{Manoj Gopalakrishnan}
 \altaffiliation[Present
address: ]{Harish-Chandra Research Institute, Chhatnag Road, Jhunsi, Allahabad
211019, India}
 \email{manoj@mri.ernet.in}
 \author{Peter Borowski}
\altaffiliation[Present address: ]{Department of Mathematics, The University of British Columbia, 1984 Mathematics Road, Vancouver, BC V6T 1Z2, Canada}
\author{Frank J\"ulicher}
 \author{Martin Zapotocky}
 
 \affiliation{Max
Planck Institute for the Physics of Complex Systems, N\"othnitzer Str.38,
01187 Dresden, Germany}
 \date{\today}
 
\begin{abstract}
We study the stochastic kinetics of a signaling module consisting of a
two-state stochastic point process with negative feedback. In the active
state, a product is synthesized which increases the active-to-inactive
transition rate of the process. We analyze this simple auto-regulatory module
using a path-integral technique based on the temporal statistics of
state-flips of the process. We develop a systematic framework to calculate
averages, auto-correlations, and response functions by treating the feedback
as a weak perturbation. Explicit analytical results are obtained to first
order in the feedback strength. Monte Carlo simulations are performed to test
the analytical results in the weak feedback limit and to investigate the
strong feedback regime. We conclude by relating some of our results 
to experimental observations in the olfactory and visual sensory systems.
\end{abstract}

 
 \maketitle

%%%%%%%%%%%%%%%%%%%%%%%%%%%  Introduction  %%%%%%%%%%%%%%%%%%%%%%%%%%%  

\section{Introduction}
\label{sec:introduction}

 Signal  transduction in biological cells refers to the set of processes by
which the cells receive and process information from the environment. From the
realization of the extra-cellular stimulus to its final recognition by the
cell/organism, there is, in general, a complex biochemical reaction network,
commonly referred to as the signal transduction pathway. Faithful transmission
of information through the pathway may require amplification and adaptation,
which is often accomplished through positive (amplification) and negative
(adaptation) feedback mechanisms built into the reaction network.
 
External and intrinsic noise can potentially limit the faithful transduction
of a `signal' (information encoded in the intensity of the external stimulus
and its spatio-temporal variation)~\cite{Berg_1977,Bialek_review}. Since
cellular biochemical networks often function with a limited number of reacting
molecules, noise is always present, and an important question is how the cell
physiology maintains its robustness in spite of the randomness in the
underlying molecular events~\cite{RAO_ARKIN}. In the context of signal
transduction, external noise refers to the randomness in the signal
(fluctuations in the local concentration of the extracellular ligand etc.),
whereas intrinsic noise is a collective effect produced by the inherent
stochasticity of the transduction mechanisms itself, such as random opening
and closing of the ion channels, fluctuations in reaction rates and so on
\cite{Swain:2002de,Kepler:2001,MAZZAG,Paulsson:2004,Bhalla:2004,Falcke:2000,Bialek_2005,Goychuk_2005}. 
Although randomness and noise are beneficial in some instances (population
heterogeneity being a well-known example~\cite{Arkin:1998ig}), by and large,
cells have evolved the means to control biochemical noise through regulatory
mechanisms (such as negative feedback) to ensure the robustness of biochemical
networks~\cite{Vilar:2002jg}. Indeed, in the context of electrical circuits, 
negative feedback is a well-known 
noise-reduction mechanism used in devices such as amplifiers and oscillators (see \cite{IEEE} for a review and important references).

Complex signaling pathways can often be
productively viewed as consisting of recurring {\it
modules}~\cite{LEIBLER,Yeger_2004}. A module is typically made up of multiple
species of interacting molecules acting together with a specific function. A
signal transduction pathway may thus be described in terms of a series of
modules, interacting with each other~\cite{DETWILER}. In very general terms, a
module receives a signal (from the extracellular environment, or from another
module) and transmits it, possibly in a different form (e.g. chemical signals
converted to electrical pulses). Each module can function to a certain extent
in isolation, but interaction among modules and their coordination are crucial
for carrying out all the complex functions of the cell.
 
\begin{figure}[h!]
\psfrag{Rm(c)}{$R_-(c)$}
 \psfrag{Rp}{$R_+$}
 \psfrag{l}{$\bar{\lambda}$}
\psfrag{J}{$J$}
 \psfrag{A*}{A*}
 \psfrag{A}{A}
 \psfrag{C}{C}
\includegraphics[width=.45\columnwidth]{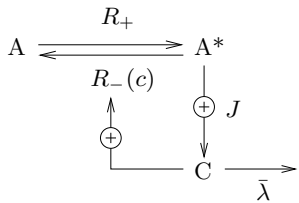}
\caption{The simple two-component auto-regulatory signaling module with
negative feedback studied in this paper. The product C (with concentration
$c$), generated with rate $J$ in the active state A*, enhances the
deactivation rate $R_-$. C is removed with a rate $\bar{\lambda}$ independent
of the activation state of A. An input signal encoded in a temporal change
$R_+(t)$ is transduced into an output signal $c(t)$.}
\label{fig:fig2}
\end{figure}

\begin{figure}[h!]
\includegraphics[width=.75\columnwidth]{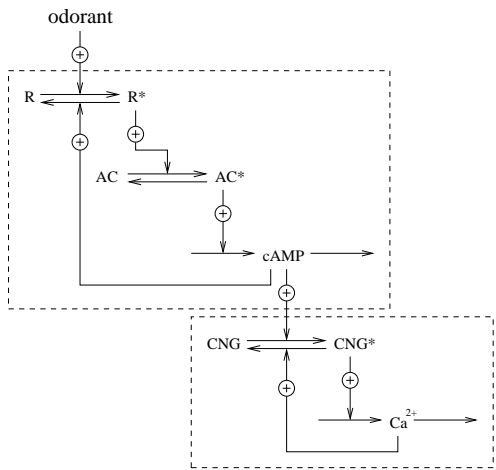}
\caption{A schematic representation of the modularity of signal transduction
in olfactory sensory neurons. (The arrows with + sign indicate enhancement of
the reaction rates, arrows without signs indicate the direction of the
reactions in the conventional sense.) After binding of an odorant (external
stimulus) to the receptor R, the enzyme adenylate cyclase (AC) is activated and
produces the 'second messenger' cAMP, leading in turn to the opening of
cyclic-nucleotide-gated (CNG) ion channels and to the depolarization of the
cell membrane.   Two negative feedback mechanisms (reviewed in~\cite{ZUFALL})
are shown in the figure: (i) cAMP-activated kinase phosphorylates (and
therefore deactivates) the receptor R, (ii) Ca$^{2+}$ ions bound to calmodulin
inhibit the Ca$^{2+}$ channels, which is crucial for adaptation to repeated
odorant stimuli~\cite{Bradley,REIDL2005}. The signal transduction pathway can
be viewed as consisting of two modules of the type shown in
Fig.~\ref{fig:fig2}, connected in series. }
\label{fig:fig1}
\end{figure}

 In this paper, we
analyze a signaling module based on a single protein that switches between
active (A*) and inactive (A) states. In the active state, a certain molecular
species C is produced with a fixed rate. The accumulation of C increases the
A*$\rightarrow$A transition rate (Fig.~\ref{fig:fig2}), leading to negative
feedback on the production of C. In addition, C is removed at a fixed rate
independent of the activation state. 
As an example, Fig.~\ref{fig:fig1} illustrates the modularity of signal
transduction and the occurrence of the module defined in Fig.~\ref{fig:fig2},
for the specific case of vertebrate olfactory sensory neurons (reviewed
in~\cite{FIRESTEIN, SCHILD}). (A closely related pathway also operates in cone
photoreceptors in the retina.)
 
 In the work presented, we develop a systematic analytical framework to
study the dynamics of the simple auto-regulatory module shown in
Fig.~\ref{fig:fig2}. Without the feedback, the activation and deactivation
kinetics would be described by a two-state Poisson process. In the presence of
feedback, the deactivation rate, at any given time, becomes dependent on the
history of transitions, and makes the effective two-state kinetics
non-Markovian. We use a path-integral technique to evaluate the effects of
feedback perturbatively, which enables us to derive the noise from the
dynamics of the system. This is in contrast to stochastic analysis of
biochemical networks based on Langevin equations~\cite{VAN_KAMPEN} with
additive white noise (see, e.g.~\cite{DETWILER,THATTAI}).
 
 We obtain
explicit analytical expressions for the different statistical averages as well
as correlation and response functions of the module in the limit of weak
feedback. We also perform Monte Carlo simulations to test the analytical
results in the weak feedback limit and to investigate the strong feedback
regime. The path-integral formalism we develop is similar to the one described
in~\cite{GOYCHUK}, and the system considered is the same as
in~\cite{MAZZAG}. The calculations in~\cite{MAZZAG}, however, focus on steady
state probability distributions rather than temporal correlations.
 
 The
paper is arranged into four sections. Sec.~\ref{sec:model} provides a brief
description of our simple model of a stochastic signaling module with
feedback. All important quantities are introduced and the notation is
defined. In Sec.~\ref{sec:path-integral}, we present and develop the
path-integral formalism. In Sec.~\ref{sec:Greensfunction}, we compute the
different Green's functions using this formalism to first order in feedback
strength. Sec.~\ref{sec:averages} is devoted to the calculation of the
averages and correlation functions, and Sec.~\ref{sec:responsefunctions}
contains the calculation of the response functions. Sec.~\ref{sec:numerics}
presents the results of Monte Carlo simulations and their comparison to
analytical results. Our conclusions and outlook for further extensions are
presented in Sec.~\ref{sec:conclusions}. Four appendices detail some of the
analytical calculations.

%%%%%%%%%%%%%%%%%%%%%%%%%%%  The model  %%%%%%%%%%%%%%%%%%%%%%%%%%%  

\section{Two-state model: General Set-up}
\label{sec:model}

\begin{figure}[h]
\psfrag{Ca}{Ca$^{2+}$}
 \psfrag{V}{$V$}
 \psfrag{l}{$\bar{\lambda}$}
\psfrag{J}{$J$}
 \includegraphics[width=.5\columnwidth]{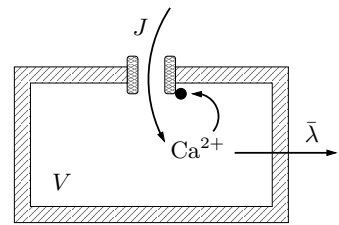}
\caption{A channel in the open state permits the entry of ions (rate $J$) into
a small compartment of volume $V$. The ions, once inside, increase the closing
rate of the channel, and are also removed from the compartment at a constant
rate $\bar{\lambda}$.}
\label{fig:fig2a}
\end{figure}

 We describe our model of a two-state signaling module using the example of
an ion channel connected to a small cellular compartment
(Fig.~\ref{fig:fig2a}).
The channel on the cell membrane exists in one of two conformational states:
{\it open}, when the ions enter through the channel, and {\it closed} when
this is not allowed. The state of the channel at time $t$ is represented by
$S(t)$, with $S=1$ being the open state and $S=0$, the closed state. The $0\to
1$ transition takes place with a rate $R_+$, and the reverse transition $1\to
0$ occurs with a rate $R_-$. Let us denote by $\bar{c}(\bar{t})$, the ion
(which in the case of the olfactory signaling pathway is Ca$^{2+}$)
concentration inside the compartment at time ${\bar t}$. The kinetics of
$\bar{c}$ is described by the equation
\begin{equation}
\frac{d\bar{c}}{d\bar{t}}=\frac{J}{V} S(\bar{t})-\bar{\lambda}
\bar{c}(\bar{t}).
\label{eq:EQ5}
\end{equation}

 In the above equation, $J$ represents the molar current of ions entering
the cell through the channel in the open state, $\bar{\lambda}$ represents the
total rate of removal of ions by membrane pumps, and $V$ is the volume of the
compartment. The channel kinetics is specified by the opening rate $R_+$ and
closing rate $R_-$. We assume the existence of a negative feedback
corresponding to a rate $R_-$ that increases with increasing ion
concentration. If the effect of the feedback is weak, one may expand to linear
order and write 
\begin{equation}
R_{-}(\bar{c})\approx R_{-}^{0}+\bar{\alpha} \bar{c}(\bar{t}),
\label{eq:EQ4a}
\end{equation}   
where $R_{-}^{0}$ is the closing rate of the channel when no ions are present
in the compartment and the coupling parameter $\bar{\alpha}$ specifies the
feedback strength. We further assume that, in general, an external signal is
received by the module as a change in the opening rate (due, e.g., to the
binding of extracellular ligands to the channel). We may write 
\begin{equation}
R_{+}(\bar{\phi}) = R_{+}^{0}+\bar{\phi}(\bar{t}),
\label{eq:EQ4b}
\end{equation}                                                                                      
where $R_{+}^{0}$ is the intrinsic opening rate of the channel in the absence
of external stimulus, and $\bar{\phi}(\bar{t})$ defines the
stimulus. Eqs.~\ref{eq:EQ5},~\ref{eq:EQ4a} and~\ref{eq:EQ4b} specify the
dynamics of the problem.
 
\begin{figure}[h!]
\psfrag{T}[cc]{$\quad t$}
 \psfrag{S(T)}{$S(t)$}
\psfrag{c(T)(l=5.)}{$c(t)$}
 \psfrag{c(T)(l=.5)}{$c(t)$}
\includegraphics[width=.75\columnwidth]{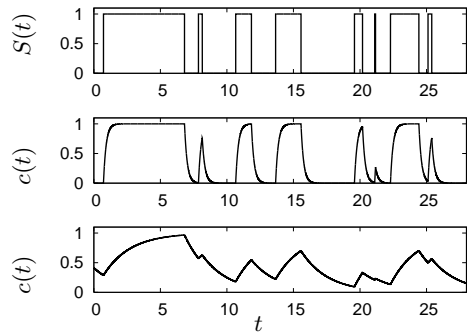}
\caption{The figure illustrates the kinetics of channel openings and Ca$^{2+}$
concentration for two typical runs, with $r_{+}=0.5$ and $\lambda=5$ (top and
middle) and  $\lambda=0.5$ (top and bottom) in the absence of feedback
($\alpha=0$).  Note that for larger $\lambda$, $c(t)$ almost follows the
step-like kinetics of $S(t)$.}
\label{fig:fig3}
\end{figure}

 It is now convenient to adopt a dimensionless
formulation of the problem. The inverse of the intrinsic closing rate
$R_{-}^0$ is chosen as the unit of time and the ratio $J/({\bar \lambda} V)$
(i.e. the maximum achievable ion concentration) is the unit of
$\bar{c}$. Using these two quantities, we scale all the other parameters, and
the complete list of dimensionless parameters is given below: 
\begin{eqnarray}
c=\frac{\bar{\lambda}V}{J}\bar{c};\quad \alpha=\frac{\bar{\alpha}
J/V}{\bar{\lambda} R_{-}^{0}};\quad
\lambda=\frac{\bar{\lambda}}{R_{-}^{0}};\quad
\phi=\frac{\bar{\phi}}{R_{-}^{0}};\nonumber\\
r_{+}=\frac{R_{+}}{R_{-}^{0}};\quad r_{-}=\frac{R_{-}}{R_{-}^0}=1+\alpha
c;\quad t=R_{-}^{0}\bar{t}.
\label{eq:EQ6}
\end{eqnarray}
The dynamical equation for $c(t)$ is then simplified to
$\frac{dc}{dt}=\lambda(S(t)-{c(t)})$, with the solution
\begin{equation}
c(t)=\lambda\int_{-\infty}^{t}e^{-\lambda(t-t^{\prime})}S(t^{\prime})dt^{\prime}.
\label{eq:EQ8}
\end{equation}
Fig.~\ref{fig:fig3} shows two typical time evolution curves of $S(t)$ and
$c(t)$. Note, from Eq.~\ref{eq:EQ8}, that in the limit $\lambda\to\infty$,
$c(t)\approx S(t)$, which is illustrated in Fig.~\ref{fig:fig3}. In this
limit, the Ca$^{2+}$ is drained from the compartment as soon as the channel
closes, and consequently, the time evolution of $c(t)$ closely follows the
channel state. In this sense, $\lambda$ is an important control parameter in
our model, and determines how much the dynamic characteristics of $c(t)$
(including fluctuations and response functions) are tied to the corresponding
quantities for the channel state. These aspects will be discussed more in
Sec.~\ref{sec:averages} and later.  

 The combined set of Eqs.~\ref{eq:EQ5}-\ref{eq:EQ4b} describe the kinetics
of $S(t)$, which may be characterized using, for instance, the $n$-point
functions ($n=1,2,...$): $\langle S(t_0)...S(t_{n-1})\rangle$. The
corresponding functions for $c(t)$ are then easily computed from these using
Eq.~\ref{eq:EQ8}:
\begin{eqnarray}
\langle c(t_0)...c(t_{n-1})\rangle=\lambda^n e^{-\lambda(t_0+\cdots
+t_{n-1})}\times\nonumber\\
\int_{-\infty}^{t_0}dt'_0...\int_{-\infty}^{t_{n-1}}dt'_{n-1}
e^{\lambda(t'_0+\cdots +t'_{n-1})}\langle S(t'_0)...S(t'_{n-1})\rangle.
\label{eq:EQ9}
\end{eqnarray}
The angular brackets $\langle\cdots\rangle$ represent statistical averaging
over the different temporal histories of the process. In particular, when the
external signal is time-independent, the system reaches a steady state, where
the one-point functions $\langle S\rangle$ and $\langle c\rangle$ are
constants, while the auto-correlation and cross-correlation functions 
\begin{eqnarray}
{\cal C}_{S}(t)=\lim_{t_0\to\infty}\langle S(t_0)S(t_0+t)\rangle-\langle
S\rangle^2, \nonumber\\ 
 {\cal C}_{c}(t)=\lim_{t_0\to\infty}\langle
c(t_0)c(t_0+t)\rangle-\langle c\rangle^2, \nonumber\\
 {\cal
C}_{Sc}(t)=\lim_{t_0\to\infty}\langle S(t_0)c(t_0+t)\rangle-\langle
S\rangle\langle c\rangle,
\label{eq:EQ8++}
\end{eqnarray}
are stationary in time. In particular it also follows from Eq.~\ref{eq:EQ8}
that
\begin{equation}
\langle c\rangle=\langle S\rangle
\label{eq:EQ9+}
\end{equation}
in the steady state. We may also define the power spectral densities for the
fluctuations in $S$ and $c$ from the stationary auto-correlation functions,
for example,
\begin{eqnarray}
P_{c}(\omega)=2\int_{0}^{\infty}{\cal C}_{c}(t)\cos(\omega
t)dt=\frac{\lambda^2}{\lambda^2+\omega^2}P_{S}(\omega),
\label{eq:EQ9+-}
\end{eqnarray}
where the relation between $P_{c}$ and $P_{S}$ follows from
Eq.~\ref{eq:EQ8}. Note that for $\omega\ll\lambda$, $P_c \simeq P_{S}$,
whereas when $\omega\gg\lambda$, $P_{c}\sim \omega^{-2}P_{S}$. This has an
interesting physical interpretation: when $\lambda$ is small over the time
scales of interest, Ca$^{2+}$ kinetics is slow, and this effectively
suppresses the power-spectrum at larger frequencies compared to the case of
large $\lambda$.  
 
 The response of the system to a time-dependent
external perturbation $\phi(t)$ is $\langle S(t)\rangle^{\phi}-\langle
S\rangle$ and $\langle c(t)\rangle^{\phi}-\langle c\rangle$, where the
superscript $\phi$ indicates that the function needs to be evaluated in the
presence of the perturbation. When the external signal is weak,
i.e. $\phi(t)\ll r_{+}$, the response is characterized using the linear
response functions $\chi_{S}(t)$ and $\chi_{c}(t)$ defined through the
following relations:
\begin{align}
\chi_S(t)=&\frac{d}{dt}\lim_{\phi\to 0}\frac{1}{\phi}\left(\langle
S\rangle^{\phi}-\langle S\rangle\right), \nonumber \\
\chi_c(t)=&\frac{d}{dt}\lim_{\phi\to 0}\frac{1}{\phi}\left(\langle
c\rangle^{\phi}-\langle c\rangle\right).
\label{eq:EQ9+++}
\end{align}
Using Eq.~\ref{eq:EQ8}, the following relation between the response functions
is also easily proved:
\begin{equation}
\chi_{c}(t)=\lambda\int_{0}^{t}\chi_{S}(t^{\prime})e^{-\lambda(t-t^{\prime})}dt^{\prime}.
\label{eq:EQ9++++}
\end{equation}

In the following sections, we present explicit calculations of the correlation and linear response functions of the model in the presence of feedback. Most of the specific results in this paper are derived for a constant $r_+$.

%%%%%%%%%%%%%%%%%%%%%%%%%%%  Path-integral description  %%%%%%%%%%%%%%%%%%%%%%%%%%%  

\section{Path-integral formulation}
\label{sec:path-integral}

 We start by introducing the propagator $\Pi_{ij}(t_0,c_0;t,c)$ with
$i,j=\{0,1\}$ that gives the probability density to find $S(t)=j$ and $c(t)=c$
given that $S(t_0)=i$ and $c(t_0)=c_0$. The mean open fraction of the channel
in steady state can be written as
\begin{equation}
\langle S\rangle =\int_0^1 \Pi_{i1}(-\infty,c_0;0,c)dc
\label{eq:<S>}
\end{equation}
and is independent of $i$ and $c_0$. We define two different mean values of
the concentration $c$, namely the mean concentrations during periods when the
channel is closed and open, respectively, starting from a special initial
state $S(t=0)=0$, $c(t=0)=0$:
\begin{align}
\langle c(t)\rangle_0 = & \int_0^1 \Pi_{i0}(0,0;t,c)c \,dc,\nonumber \\
\langle c(t)\rangle_1 = & \int_0^1 \Pi_{i1}(0,0;t,c)c \,dc.
\label{eq:<c>_01}
\end{align}

 The auto-correlation function of $S$ in steady state is
\begin{equation}
\langle S(0)S(t)\rangle = \int_0^1\!dc\int_0^1\!dc_1
\Pi_{i1}(-\infty,c_0;0,c_1)\Pi_{11}(0,c_1;t,c).
\label{eq:<S(0)S(t)>}
\end{equation}
The linear response function can be written using a step function
$r_+\rightarrow r_++\phi(t)$ (with $\phi(t)=\phi_0\theta(t)$) as a stimulus:
\begin{align}
\chi_S(t)= & \frac{\partial}{\partial t}\left.\frac{\partial}{\partial
 \phi_0}\right|_{\phi_0=0}\sum_j\int_0^1\!dc\int_0^1\!dc_1\Pi_{ij}(-\infty,c_0;0,c_1)\times
 \nonumber \\
 & \qquad \Pi_{j1}^{\phi_0}(0,c_1;t,c).
\label{eq:chi_Pi}
\end{align}

 In order to formulate a path-integral representation of the propagator, we
define the functional ${\cal P}_{ij}[t_0,t;\{\tau_k\}_{k=1}^{N};c_0;N]$ of
$S(t)$ with $t>t_0$. 
 It represents the probability that a certain time
evolution of the system from 
 $S(t_0)=i$ and $c(t_0)=c_0$ to $S(t)=j$ is
realized. This time evolution is 
 characterized by the set of times
$\{\tau_k\}_{k=1}^{N}$ at which the state 
 switches from $S=0$ to $1$ or
from $S=1$ to $0$. The number $N$ 
 represents the total number of such state
changes in the time interval $[t_0,t]$ 
 and is even if $i=j$ and odd
otherwise. The propagator $\Pi_{ij}(t_0,c_0;t,c)$ 
 can be expressed as a sum
over all possible realizations 
of
$S(t)$ as:
\begin{equation}
\Pi_{ij}(t_0,c_0;t,c)=\sum_{N}\int {\cal D\tau}{\cal
P}_{ij}[t_0,t;\{\tau_k\}_{k=1}^{N};c_0;N]\delta[c(t)-c],
\end{equation}
where $\int {\cal D}\tau\equiv
\int_{t_0}^{t}d\tau_1\int_{\tau_1}^{t}d\tau_2...\int_{\tau_{N-1}}^{t}d\tau_N$.

 It is convenient to introduce the reduced propagator or Green's function
\begin{equation}
G_{ij}(t_0,c_0;t)\equiv \int_0^1\Pi_{ij}(t_0,c_0;t,c)dc,
\label{eq:reduced_propagator}
\end{equation}
which can be written as
\begin{equation}
G_{ij}(t_0,c_0;t)=\sum_{N}\int {\cal D\tau}{\cal
P}_{ij}[t_0,t;\{\tau_k\}_{k=1}^{N};c_0;N].
\label{eq:EQ10}
\end{equation}
By definition the following relations hold:
\begin{equation}
G_{01}+G_{00}=1=G_{10}+G_{11}.
\label{eq:EQ10+}
\end{equation}

 Note that in the absence of feedback, the functionals ${\cal P}_{01}$ and
${\cal P}_{11}$ do not depend on $c$, and the history-dependence is thus
absent. In this case, the expression for the $n$-point functions reduces to
the product of Green's functions: $\langle
S(t_0)...S(t_{n-1})\rangle=G_{01}^{(0)}(-\infty,t_0)G_{11}^{(0)}(t_0,t_1)\cdots
G_{11}^{(0)}(t_{n-2},t_{n-1})$, where the superscript $(0)$ denotes the
absence of feedback (valid for the rest of the paper and all quantities)
\footnote{Higher order correlation functions can be expressed in the path
integral representation as $\langle
S(t_0)...S(t_{n-1})\rangle=\sum_{m_1,...,m_n=0}^\infty \int {\cal
D}{\{\tau\}_1}{\cal P}_{01}[-\infty,t_0;\{\tau\}_1;0;2m_1+1]\times \\ \int
{\cal D}{\{\tau\}_2}{\cal P}_{11}[t_0,t_1;\{\tau\}_2;c_0;2m_2]\cdots \times \\
\int {\cal D}{\{\tau\}_n}{\cal
P}_{11}[t_{n-2},t_{n-1};\{\tau\}_n;c_{n-2};2m_n]$. In the presence of a
feedback, the weights ${\cal P}_{ij}[t_0,t;\{\tau_k\}_{k=1}^N;c_0;N]$ are
functions of the concentration $c_0$ at time $t_0$. Since for any realization
$S(t)$, the function $c(t)$ follows according to Eq. \ref{eq:EQ8}, the values
$c_k=c(t_{k})$ are nontrivial. As a consequence, correlation functions of
order larger than two cannot be written as simple products of propagators.}.
As special cases, the one- and two-point functions in the absence of feedback
are given by
\begin{eqnarray}
\langle S(t_0)\rangle & = & \langle S\rangle=G_{01}^{(0)}(-\infty,t_0),
\nonumber\\
 \langle S(t_0)S(t)\rangle & = &
G_{01}^{(0)}(-\infty,t_0)G_{11}^{(0)}(t_0,t),
\label{eq:EQ11++}
\end{eqnarray}
when $\alpha=0$.
 
 We now write down the functionals ${\cal P}_{ij}$ in the
explicit form. The expressions are presented only for ${\cal P}_{00}$,
extension to the other cases are straightforward. Furthermore, since the
different Green's functions are related (Eq.~\ref{eq:EQ10+} and
Eq.~\ref{eq:EQ27} later), it is enough to compute one of them. It is
instructive to start with $\alpha=0$, where the transitions $0\to 1$ and $1\to
0$ are characterized by the (time-invariant) rates $r_+$ and
$r_{-}=r_{-}^{0}\equiv 1$, respectively (in terms of scaled time). Let us
consider a certain history of the process such that the channel is closed at
$t=0$ and $t$, but changes state an even number $N$ times in between, at times
$\tau_1,\tau_2,...,\tau_N$. First consider the interval $[0,\tau_1]$. The
probability that the channel remains in state 0 until time $\tau_1$ has the
form $P_{0}(0,\tau_1)=e^{-r_+\tau_1}$. Similarly, the probability of the
channel to remain in state 1 during the time interval $[\tau_1,\tau_2]$ is
$P_{1}(\tau_1,\tau_2)=e^{-(\tau_2-\tau_1)}$. The (differential) probability
for the whole process is
\begin{eqnarray}
{\cal
  P}_{00}^{(0)}[0,t;\{\tau_{i}\}_{i=1}^{N};N]d\tau_1d\tau_2...d\tau_N=r_{+}^{\frac{N}{2}}\times\nonumber\\
  \prod_{i=1}^{N}d\tau_iP_{0}(0,\tau_1)P_{1}(\tau_1,\tau_2)P_{0}(\tau_2,\tau_3)\cdots
  P_{0}(\tau_N,t).
\label{eq:EQA3}
\end{eqnarray}

\begin{figure}[h!]
\psfrag{tau1}[cc]{$\tau_1$}
 \psfrag{tau2}[cc]{$\tau_2$}
\psfrag{tau3}[cc]{$\tau_3$}
 \psfrag{tau4}[cc]{$\tau_4$}
\psfrag{tau2m-1}[cc]{$\tau_{2m-1}$}
 \psfrag{tau2m}[cc]{$\tau_{2m}$}
\psfrag{t1}{$t_1$}
 \psfrag{t2}{$t_2$}
 \psfrag{tm}{$t_m$}
\psfrag{t1p}{$t'_1$}
 \psfrag{t2p}{$t'_2$}
 \psfrag{tmp}{$t'_m$}
\psfrag{t}[cc]{$t$}
 \psfrag{S}{$S$}
 \psfrag{1}{1}
 \psfrag{0}[cc]{0}
\psfrag{y0}{0}
 \includegraphics[width=.85\columnwidth]{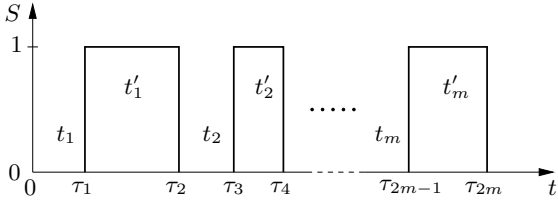}
\caption{A schematic diagram of time evolution of $S(t)$, showing the time
flip variables $\tau_i$ versus the interval variables $t_i$ and
$t^{\prime}_i$.}
\label{fig:fig4}
\end{figure}

 We now substitute for $P_0$ and $P_1$, and make the transformation from
flip-times $\{\tau_i\}_{i=1}^{N}$ to time-interval variables
$\{t_{j},t_{j}^{\prime}\}_{j=1}^{m}$ with $m=N/2$, where the $\{t_j\}$ denote
the times, when the channel is closed and the $\{t'_j\}$ when it is open (see
Fig.~\ref{fig:fig4}). The relation between the flip-times and the time
intervals is summarized below:
\begin{align}
\tau_1 & = t_1, & & \nonumber\\
 \tau_i &
=\sum_{l=1}^{i/2}t_{l}+\sum_{l=1}^{i/2}t^{\prime}_{l} \qquad \mathrm{even}
\quad i\geq 2, \nonumber\\
 \tau_j &
=\sum_{l=1}^{(j+1)/2}t_{l}+\sum_{l=1}^{(j-1)/2}t^{\prime}_{l} \qquad
\mathrm{odd} \quad j>1.
\label{eq:EQA3+}
\end{align}

 The Jacobian for the transformation is $J=1$, as follows from
Eq.~\ref{eq:EQA3+}. The probability functional thus has the form
\begin{equation}
{\cal
P}_{00}^{(0)}[0,t;\{t_i\},\{t^{\prime}_i\};2m]=r_{+}^me^{-F_{00}(0,t;\{t^{\prime}_i\};m)},
\label{eq:EQ13}
\end{equation}
with the weight factor $F_{00}$ given by
\begin{equation}
F_{00}(0,t;\{t^{\prime}_i\};m)=r_{+}t+(1-r_+)\sum_{i=1}^{m}t^{\prime}_i.
\label{eq:EQ15}
\end{equation}

 In the presence of the feedback-coupling to $c(t)$, the expression for
${\cal P}_{00}$ is modified to
\begin{eqnarray}
{\cal
P}_{00}[0,t;\{t_i\},\{t^{\prime}_i\};c_0;2m]=r_{+}^{m}\times\nonumber\\
\sideset{}{'}\prod_{i=2}^{2m} \left(1+\alpha c(\tau_i)\right)
e^{-F_{00}-\alpha\sideset{}{''}\sum_{j=1}^{2m-1}\int_{\tau_j}^{\tau_{j+1}}d\tau
c(\tau)},
\label{eq:EQ16}
\end{eqnarray}
where the prime (double prime) on the product- or sum-symbol indicates that
the running index is always even (odd). The above expression may be expanded
in powers of the dimensionless feedback strength $\alpha$ as follows:
\begin{eqnarray}
{\cal
P}_{00}[0,t;\{t_i\},\{t^{\prime}_{i}\};c_0;2m]=r_{+}^{m}e^{-F_{00}}\bigg[1+\alpha\times\nonumber\\
\left(
\sideset{}{'}\sum_{i=2}^{2m}c(\tau_i)-\sideset{}{''}\sum_{j=1}^{2m-1}\int_{\tau_{j}}^{\tau_{j+1}}d\tau
c(\tau) \right)\bigg] +O(\alpha^2).
\label{eq:EQ17}
\end{eqnarray}
Eq.~\ref{eq:EQ17} is the basis for more specific and detailed calculations to
follow in the next sections.

%%%%%%%%%%%%%%%%%%%%%%%%%%%%%%%%%%%%%%%%%%%%%%%%%%%%%%%%%%%%%%%%%%%%%%%%%%%

\section{Computation of Green's functions}
\label{sec:Greensfunction}

\subsection{Computation of $\boldsymbol{G_{00}}$}

 For instructive purposes, we first show how the Green's function
$G_{00}^{(0)}(0,t)$ in the absence of feedback is computed using our
formalism. In this case the exact answer is easily found  by solving its rate
equation
\begin{equation}
\partial_{t}G_{00}^{(0)}=r_{-}-(r_{-}+r_+)G_{00}^{(0)},
\label{eq:EQ17+}
\end{equation}
with $r_{-}=r_-^0=1$ and the initial condition $G_{00}^{(0)}(0,0)=1$. The
solution is
\begin{equation}
G_{00}^{(0)}(0,t)=\frac{1}{1+r_+}\left(1+r_+e^{-(1+r_+)t}\right).
\label{eq:EQA5}
\end{equation}

 This Green's function can also be computed using the path-integral
technique outlined previously, more specifically from Eq.~\ref{eq:EQ10} using
Eqs.~\ref{eq:EQ13} and~\ref{eq:EQ15}. For simplicity, let us put $t_{0}=0$,
and define the Laplace transform of the Green's function ${\tilde
G}_{00}^{(0)}(s)=\int_{0}^{\infty}G_{00}^{(0)}(0,t)e^{-st}dt$. The calculation
is most easily done using the generalized convolution theorem presented in
Appendix A (Eq.~\ref{eq:EQB0-}). Using this technique, the path integral in
Eq.~\ref{eq:EQ10} becomes a product in the $s$-space, and the result is
\begin{align}
{\tilde G}_{00}^{(0)}(s)= & g(s+r_+), \nonumber \\
 \mathrm{with}\quad g(s)=
&
\frac{1}{s}\sum_{m=0}^{\infty}\left(\frac{1}{s}\right)^{m}\left(\frac{r_+}{s+1-r_+}\right)^{m}.
\label{eq:EQA4}
\end{align}
After summing the geometric series in Eq.~\ref{eq:EQA4}, we find that ${\tilde
G}_{00}^{(0)}(s)=(s+1)/\left[s(s+1+r_+)\right]$, which, upon inversion, gives
Eq.~\ref{eq:EQA5}.
 
 Let us now extend the previous calculation to include
feedback, and compute $G_{00}(0,c_0;t)$ to first order in $\alpha$:
\begin{equation}
G_{00}(0,c_0;t) = G_{00}^{(0)}(0,t) + \alpha G_{00}^{(1)}(0,c_0;t) +
O(\alpha^2).
\end{equation}
From Eq.~\ref{eq:EQ17} and Eq.~\ref{eq:EQ10}, we find that
$G_{00}^{(1)}(0,c_0;t)$ depends only linearly on $c_0$: $G_{00}^{(1)}(0,c_0;t)
= G_{00}^{(1)}(0,c_0=0;t) + c_0f(t)$. Putting together the $c_0$-independent
terms, we write
\begin{equation}
G_{00}(0,c_0;t)=G_{00}(0,c_0=0;t)+\alpha c_0f(t)+ O(\alpha^2),
\label{eq:EQ18}
\end{equation}
where $c_{0}f(t)$ is simply the $O(\alpha)$ term in $G_{00}(0,c_0;t)$, when
the channel evolves in a situation with $J\equiv 0$ so that $r_{-}(t)=1+\alpha
c_{0}e^{-\lambda t}$. Using this $r_-(t)$, the function $f(t)$ is easily
determined by solving the rate equation Eq.~\ref{eq:EQ17+} with the initial
condition $G_{00}(0,c_0;0)=1$. The result is
\begin{eqnarray}
f(t)=\frac{r_+}{\lambda}\Bigg(\frac{-1}{1+r_+-\lambda}e^{-(1+r_+)t}+\frac{1}{1+r_+}e^{-(1+r_++\lambda)t}
 \nonumber \\
 +\frac{\lambda}{(1+r_+)(1+r_+-\lambda)}e^{-\lambda
 t}\Bigg).~~~~~
\label{eq:EQ19}
\end{eqnarray}

 The function $G_{00}(0,c_0=0;t)$ itself may be expressed in terms of three
integrals as follows:
\begin{eqnarray}
G_{00}(0,c_0=0;t)= e^{-r_{+}t}\Bigg( 1+\sum_{m=1}^{\infty} r_{+}^{m}\big[
 I_{0}(t;m) \nonumber\\
 + \alpha \left( I_{1}(t;m)-I_{2}(t;m)\right)\big]
 \Bigg) + O(\alpha^2),
\label{eq:EQ20}
\end{eqnarray}
where
\begin{align}
I_{0}(t;m)= & \int{\cal D}Te^{-(1-r_+)\sum_{i=1}^{m}t^{\prime}_i}, \nonumber
\\
 I_{1}(t;m)= & \int{\cal
D}Te^{-(1-r_+)\sum_{i=1}^{m}t^{\prime}_i}\sideset{}{'}\sum_{i=2}^{2m}c(\tau_i),
\nonumber \\
 I_{2}(t;m)= & \int{\cal
D}Te^{-(1-r_+)\sum_{i=1}^{m}t^{\prime}_i}\sideset{}{''}\sum_{j=1}^{2m-1}\int_{\tau_j}^{\tau_{j+1}}d\tau
c(\tau).
\label{eq:EQ21}
\end{align}
Note that we have introduced the compact notation $\int{\cal D}T\equiv
\int_{0}^{t}dt_1\int_{0}^{t-t_1}dt^{\prime}_1\int_{0}^{t-t_1-t^{\prime}_1}dt_2...\int_{0}^{t-\cdots
-t_m}dt^{\prime}_m$ for the integration measure.  The explicit calculations
are done most conveniently in terms of Laplace-transformed variables, defined
in the standard way: ${\tilde f}(s)=\int_{0}^{\infty}f(t)e^{-st}dt$. The
previous Eq.~\ref{eq:EQ20} becomes, in this notation,
\begin{eqnarray}
{\tilde G}_{00}(c_0=0;s)=(s+r_+)^{-1} \nonumber \\
+\sum_{m=1}^{\infty}r_{+}^m \left[ {\tilde I_0}(s+r_+;m) \right.\nonumber \\
\left. + \alpha\left({\tilde I_1}(s+r_+;m)-{\tilde I_2}(s+r_+;m)\right)\right]+O(\alpha^2).
\label{eq:EQ22}
\end{eqnarray}
The calculation of the integrals $\tilde{I_0}$,  $\tilde{I_1}$ and
$\tilde{I_2}$ is done in Appendix~\ref{sec:appendix_I}. We omit further
details, and present only the final result:
\begin{widetext}
\begin{eqnarray}
G_{00}(0,c_0=0;t)\equiv G_{00}^{(0)}(0,t)+\alpha G^{(1)}_{00}(0,c_0=0;t) =
 G_{00}^{(0)}(0,t) +\alpha\frac{r_+}{(1+r_+)^2}\left[
 \frac{r_++\lambda}{1+r_++\lambda}- e^{-\lambda
 t}\frac{r_{+}(1+r_+)}{(1+r_{+}-\lambda)^2} \right. \nonumber\\
 \left. -
 e^{-(1+r_++\lambda)t}\frac{1+r_+}{\lambda(1+r_++\lambda)}+
 e^{-(1+r_+)t}\left(\frac{(1+r_+)(\lambda-1)t}{1+r_+-\lambda}+\frac{r_+^2-(\lambda-1)^3+r_+(2-3\lambda+2\lambda^2)}{\lambda(1+r_+-\lambda)^2}\right)\right]
 + O(\alpha^2).
\label{eq:EQ26}
\end{eqnarray}
\end{widetext}

\subsection{Relation between $\boldsymbol{G_{11}}$ and $\boldsymbol{G_{00}}$}

 We will now show that there is a non-trivial relation between
$G_{11}(0,c_0=0;t)$ and $G_{00}(0,c_0=0;t)$, as follows:
\begin{align}
G_{11}(0,c_0=0;t)= & G_{01}(0,c_0=0;t) \nonumber \\
 & +\frac{\partial_t
 G_{00}(0,c_0=0;t)}{\partial_t G_{00}(0,c_0=0;t)|_{t=0}},
\label{eq:EQ27}
\end{align}
which will be shown to be true up to $O(\alpha)$. To prove this, let us start
with the case $\alpha=0$. Then, the following relation is true for any
arbitrary $0\leq t^{\prime}\leq t$:
\begin{equation}
G_{01}^{(0)}(0,t)=G_{00}^{(0)}(0,t^{\prime})G_{01}^{(0)}(t^{\prime},t)+G_{01}^{(0)}(0,t^{\prime})G_{11}^{(0)}(t^{\prime},t).
\label{eq:EQ28}
\end{equation}

 Let us now take the limit of $t^{\prime}\to 0$, and use the Taylor
expansions $G_{00}^{(0)}(0,t^{\prime})=1+t^{\prime}\frac{\partial
G_{00}^{(0)}(0,t)}{\partial t}|_{t=0}+\cdots$,
$G_{ij}^{(0)}(t',t)=G_{ij}^{(0)}(t-t')=G_{ij}^{(0)}(0,t)-t'\frac{\partial
G_{ij}^{(0)}(0,t)}{\partial t}+\cdots$ as well as the condition
$G_{01}^{(0)}=1-G_{00}^{(0)}$. After substituting  these back into
Eq.~\ref{eq:EQ28}, we arrive at Eq.~\ref{eq:EQ27}.
 
 In the presence of
feedback, Eq.~\ref{eq:EQ28} is not true anymore because of the explicit
history-dependence. However, using expansions of the two Green's functions
$G_{01}(t^{\prime},c^{\prime};t)$ and $G_{11}(t^{\prime},c^{\prime};t)$
similar to Eq.~\ref{eq:EQ18},
\begin{align}
G_{01}(t^{\prime},c^{\prime};t) & =G_{01}(t^{\prime},c^{\prime}=0;t)+\alpha
c^{\prime}f_1(t-t^{\prime})+ O(\alpha^2), \nonumber\\
G_{11}(t^{\prime},c^{\prime};t) & =G_{11}(t^{\prime},c^{\prime}=0;t)+\alpha
c^{\prime}f_2(t-t^{\prime})+ O(\alpha^2),
\label{eq:EQ32}
\end{align}
Eq.~\ref{eq:EQ27} can be shown to be valid also for $\alpha \ne 0$ (to
$O(\alpha)$). The proof as well as the functions $f_1(t)$ and $f_2(t)$ are
given in Appendix~\ref{sec:appendix_G11}.
 
%%%%%%%%%%%%%%%%%%%%%%%%%%%%%%%%%%%%%%%%%%%%%%%%%%%%%%%%%%%%%%%%%%%%%%%%%%%%%%%%%%%%%%%

\section{Averages and correlators}
\label{sec:averages}

\subsection{Averages and comparison to mean-field analysis}

 The mean fraction of open channels in the steady state is now easily found
as $\langle S\rangle=\lim_{t\to\infty}[1-G_{00}(0,c_0=0;t)]$, and the result
is 
\begin{align}
\langle S\rangle=\langle c\rangle= & \frac{r_+}{1+r_+}\times \nonumber \\
 &
 \left(1-\alpha \frac{r_++\lambda}{(1+r_+)(1+r_++\lambda)}\right)
 +O(\alpha^2).
\label{eq:EQ35+}
\end{align}

 For comparison, we may also compute the same quantity using the mean-field
approach. In general, the steady state obeys the relation
\begin{equation}
r_{+}(1-\langle S\rangle)=\langle r_{-} S\rangle = \langle S\rangle + \alpha
\langle c S\rangle.
\label{eq:EQ35++}
\end{equation}
In the spirit of mean-field analysis, we now assume that the fluctuations in S
and c are independent, $\langle c S\rangle = \langle c \rangle \langle
S\rangle$. Making use of the equality $\langle c\rangle=\langle S\rangle$ in
Eq.~\ref{eq:EQ35++} (valid in the steady state), one obtains a quadratic
equation for $\langle S\rangle$. The solution is
\begin{equation}
\langle S\rangle_{\mathrm{MF}}=\frac{\sqrt{(1+r_+)^2+4\alpha
r_+}-(1+r_+)}{2\alpha}.
\label{eq:EQ35+++}
\end{equation}
A small-$\alpha$ expansion of the square root gives
\begin{equation}
\langle S\rangle_{\mathrm{MF}}=\frac{r_+}{1+r_+}- \alpha
\frac{r_+^2}{(1+r_+)^3}+O(\alpha^2).
\label{eq:EQ35++++}
\end{equation}

 The mean-field treatment can be expected to be valid if the calcium
dynamics is slow compared to the channel dynamics, as in this case the calcium
concentration exhibits only small fluctuations around its mean value and can
be replaced by a constant (as will be seen explicitly in
Eq.~\ref{eq:EQ37+++-}). Indeed, the full result in Eq.~\ref{eq:EQ35+} reduces
to the mean-field result in Eq.~\ref{eq:EQ35++++} when $\lambda \ll r_{+}$ and
$\lambda \ll 1$. For all other values of $\lambda$, the mean-field analysis
underestimates the effect of feedback.
 
 A second limiting case of interest
is the situation where the pump rate is large: $\lambda\to\infty$, in which
case, from Eq.~\ref{eq:EQ8}, we find that $c(t)\approx S(t)$ at all times
$t$. In this case the closing rate may be approximated as $r_{-}\approx
1+\alpha$ since $c\simeq 1$ in the open state of the channel. The mean open
fraction, therefore, is given by
\begin{equation}
\langle S\rangle\simeq \frac{r_+}{1+r_++\alpha}\approx
\frac{r_+}{1+r_+}-\alpha\frac{r_+}{(1+r_+)^2}+O(\alpha^2).
\label{eq:EQ35+++++}
\end{equation}
It is easily verified that this expression is the limit of the one given in
Eq.~\ref{eq:EQ35+} to $O(\alpha)$ for $\lambda \gg r_{+}$ and $\lambda \gg
1$.
 
 Our perturbative result for the mean fraction of open channels
(Eq.~\ref{eq:EQ35+}) is contained in the results of~\cite{MAZZAG}.
 
\subsection{Auto-correlation functions}

 To compute the two-point function of $S$, we put the definition of the
reduced propagator (Eq.~\ref{eq:reduced_propagator}) into
Eq.~\ref{eq:<S(0)S(t)>}:
\begin{align}
\langle S(0)S(t)\rangle= & \int_0^1
dc_1\Pi_{i1}(-\infty,c_0;0,c_1)G_{11}(0,c_1;t).
\end{align}
Using Eq.~\ref{eq:EQ32} together with Eqs.~\ref{eq:<S>} and~\ref{eq:<c>_01},
we arrive at
\begin{equation}
\langle S(0)S(t)\rangle=\langle S\rangle G_{11}(0,c_0=0;t)+\alpha\langle
c\rangle_{1}f_{2}(t)+O(\alpha^2),
\label{eq:EQ36}
\end{equation}
where $G_{11}(0,c_0=0;t)$ is given by Eq.~\ref{eq:EQ27}, $G_{00}(0,c_0=0;t)$
is given by Eq.~\ref{eq:EQ26}, $f_2(t)$ is given by Eq.~\ref{eq:EQ33} and
$\langle c\rangle_{1}$ is the steady state value of the restricted average
$\langle c(t)\rangle_{1}$ (Eq.~\ref{eq:<c>_01}) and calculated in
Appendix~\ref{sec:appendix_c}. Using Eq.~\ref{eq:EQ36} and the steady state
average in Eq.~\ref{eq:EQ35+}, we find the auto-correlation function of $S$
(Eq.~\ref{eq:EQ8++}) to be
\begin{eqnarray}
{\cal C}_{S}(t)=  \frac{r_+}{(1+r_+)^2}e^{-(1+r_+)t}
 + \alpha\left(
  B_1e^{-(1+r_+)t} + \right.\nonumber\\
 \left. C_1e^{-\lambda t} +
  D_1e^{-(1+r_++\lambda)t}  + E_1te^{-(1+r_+)t}\right) + O(\alpha^2).
\label{eq:EQ37}
\end{eqnarray}
The coefficients $B_1,C_1,D_1,E_1$ are given in
Appendix~\ref{sec:coefficients}. We observe that the first order correction
term introduces two new time scales. Furthermore, the feedback introduces a
term non-monotonic in time, whose sign depends on the relative values of
$\lambda$ and $r_+$. For $\lambda<1$ and $\lambda >1+r_+$, $E_1 >0$, whereas
for intermediate values $1 <\lambda <1+r_+$, $E_1<0$. This term is the first
indication that the feedback term introduces qualitative differences in the
decay of the auto-correlation function.
 
 The corresponding correlator for
$c(t)$ is computed from Eq.~\ref{eq:EQ9} using the two-point function in
Eq.~\ref{eq:EQ36}. The result is
\begin{eqnarray}
{\cal   C}_{c}(t)=\frac{r_+\lambda}{(1+r_+)^2[(1+r_+)^2-\lambda^2]}\left.\Big(
e^{-\lambda t}(1+r_+)- \right.\nonumber \\
 \left.\lambda
e^{-(1+r_+)t}\right.\Big) + \alpha \left( B_2e^{-(1+r_+)t} + C_2e^{-\lambda t}
+ \right.\nonumber \\
 \left. D_2e^{-(1+r_++\lambda)t}+ E_2te^{-(1+r_+)t} +
F_2te^{-\lambda t} \right) + O(\alpha^2).
\label{eq:EQ37+}
\end{eqnarray}

 The coefficients $B_2,C_2,D_2,E_2,F_2$ are given in
Appendix~\ref{sec:coefficients}. Similar to the previous case, the first order
correction term has introduced two new time scales, and unlike the previous
case, there are two non-monotonic terms in time. From Eq.~\ref{eq:EQ37+} and
using Eq.~\ref{eq:EQ9+-}, we now find the power spectrum for $c$-fluctuations
to $O({\alpha})$:
\begin{eqnarray}
P_{c}(\omega)=2\frac{r_+\lambda^2}{(1+r_+)(\omega^2+\lambda^2)[\omega^2+(1+r_+)^2]}+\nonumber\\
2\alpha\left(
\frac{B_2(1+r_+)}{\omega^2+(1+r_+)^2}+\frac{C_2\lambda}{\omega^2+\lambda^2}+
\frac{D_2(1+r_++\lambda)}{\omega^2+(1+r_++\lambda)^2}+ \right.\nonumber\\
\left. \frac{E_2[(1+r_+)^2-\omega^2]}{[(1+r_+)^2+\omega^2]^2}+\frac{F_2(\lambda^2-\omega^2)}{(\lambda^2+\omega^2)^2}\right)+
O(\alpha^2).
\label{eq:EQ37+-}
\end{eqnarray}
For large $\omega$, $P_{c}(\omega)\sim \omega^{-4}$, which is true with and
without feedback~\footnote{Putting in the coefficients from
Appendix~\ref{sec:coefficients} shows that the terms $\sim \omega^{-2}$ cancel
also for $\alpha \ne 0$.}. However, for small frequencies, i.e., $\omega\ll
\lambda$ in particular, the power spectrum for $c$ effectively decays as
$\omega^{-2}$ because large $\lambda$ attenuates the high-frequency (but still
less than $\lambda$) noise.  
 
 From Appendix~\ref{sec:coefficients}, we
see that $F_{2}<0$ always, whereas $E_{2}<0$ for $1 < \lambda <1+r_+$ and
$E_{2}>0$ for $\lambda<1$ and  $\lambda >1+r_+$. The interplay of the time
scales $\lambda$ and $1+r_+$ gives rise to non-trivial time-dependence in the
auto-correlation function. This results in a variety of possible time courses,
examples of which can be observed in numerical simulations presented in
Sec.~\ref{sec:numerics}.
 
 From the correlators Eq.~\ref{eq:EQ37} and
Eq.~\ref{eq:EQ37+}, the fluctuations are also determined easily. We define the
mean-squared fluctuations through
\begin{align}
(\delta S)^2= & \langle S^2\rangle-\langle S\rangle^2=\langle
S\rangle(1-\langle S\rangle)={\cal C}_{S}(0), \nonumber \\
 (\delta c)^2= &
\langle c^2\rangle-\langle c\rangle^2={\cal C}_{c}(0).
\label{eq:EQ37++}
\end{align}

 The channel state fluctuation has the following characteristic: Since
$(\delta S)^2=\langle S\rangle(1-\langle S\rangle)$, its maximum is always
$\frac{1}{4}$, which is reached when $\langle S\rangle=\frac{1}{2}$,
independent of feedback. The RMS fluctuations, to first order in $\alpha$ are
given by
\begin{align}
\delta S= &
\frac{\sqrt{r_+}}{1+r_+}\left(1-\frac{\alpha}{2}\frac{(1-r_+)(r_++\lambda)}{(1+r_+)(1+r_++\lambda)}\right)
+ O(\alpha^2), \nonumber\\
 \delta c= & \sqrt\frac{r_+
\lambda}{(1+r_+)^2(1+r_++\lambda)} \left.\bigg( 1+\frac{\alpha}{2}\times
\right.\nonumber \\
 &
\left. \frac{r_+^3+r_+^2(3\lambda-2)+r_+(2\lambda^2-4\lambda-3)-\lambda(2\lambda+3)}{(1+r_+)(1+r_++\lambda)(1+r_++2\lambda)}\right.\bigg)
\nonumber \\
 & \qquad \qquad  + O(\alpha^2).
\label{eq:EQ37+++}
\end{align}

 For the channel state, we see that the first order correction term simply
reverses sign at the point $r_+=1$ (when the intrinsic opening and closing
rates are the same). The fluctuations are enhanced by feedback when $r_+<1$
and suppressed when $r_+>1$.
 
 The calcium fluctuation can likewise be
increased or decreased by feedback, depending on the values of $\lambda$ and
$r_+$. The precision of output of the signaling module is given by the
relative fluctuation (normalized standard deviation) in Ca$^{2+}$. Using
Eqs.~\ref{eq:EQ35+} and~\ref{eq:EQ37+++} we obtain
\begin{align}
\frac{\delta c}{\langle c\rangle}= &
 \sqrt{\frac{\lambda}{r_+(1+r_++\lambda)}}\left.\bigg(1+\alpha\times
 \right.\nonumber\\
 &
 \left. \frac{(r_++\lambda)(-1+r_++2\lambda)}{2(1+r_++2\lambda)(1+r_++\lambda)}\right.\bigg)
 + O(\alpha^2).
\label{eq:EQ37+++-}
\end{align}

 Clearly for $\lambda > \frac{1}{2}$ the relative fluctuation is increased
for any value of $r_+$, while for sufficiently small $\lambda$ it is increased
when $r_+$ is large and decreased when $r_+ <1$. The precision of output can
thus be improved by feedback when the input is weak ($r_+$ is small).

%%%%%%%%%%%%%%%%%%%%%%%%%%%%%%%%%%%%%%%%%%%%%%%%%%%%%%%%%%%%%%%%%%%%%%%%%%

\section{Response functions}
\label{sec:responsefunctions}

\subsection{Computation of $\boldsymbol{\chi_S}$ and $\boldsymbol{\chi_c}$}

 When a step-stimulus $r_++\phi_0\theta(t)$ is applied to the system, the
response function (Eq.~\ref{eq:chi_Pi}) in $S$ can be written using the
definition of the Green's function (Eq.~\ref{eq:reduced_propagator}) together
with Eq.~\ref{eq:EQ32}, as
\begin{align}
\chi_S(t) & =\frac{\partial}{\partial t}\left.\frac{\partial}{\partial
 \phi_0}\right|_{\phi_0=0}\left[\langle S\rangle G_{11}^{\phi_0}(0,0;t) +
 (1-\langle S\rangle)G_{01}^{\phi_0}(0,0;t)\right.\nonumber \\
 & \left. +
 \alpha\left( f_1^{\phi_0}(t)\langle c\rangle_0 +  f_2^{\phi_0}(t)\langle
 c\rangle_1\right)\right] + O(\alpha^2),
\end{align}
where $\phi_0$ appearing as superscript indicates that $r_+$ is to be replaced
by $r_++\phi_0$. Putting in the two Green's functions (Eqs.~\ref{eq:EQ26}
and~\ref{eq:EQ27}) together with Eqs.~\ref{eq:EQ33} and the restricted calcium
averages (computed in Appendix~\ref{sec:appendix_c}), the final result is
\begin{eqnarray}
\chi_S (t>0) = \frac{1}{1+r_+}e^{-(1+r_+)t}
 + \alpha\left.\bigg(
  B_3e^{-(1+r_+)t} + \right.\nonumber\\
 \left. C_3e^{-\lambda t} +
  D_3e^{-(1+r_++\lambda)t}  + E_3te^{-(1+r_+)t} \right.\bigg) + O(\alpha^2),
\label{eq:EQ43}
\end{eqnarray}
where $B_3,C_3,D_3,E_3$ are given in Appendix~\ref{sec:coefficients}. The
coefficient $E_3$ of the non-monotonic term is negative for $\lambda <1$ and
$\lambda>1+r_+$, and positive for $1<\lambda<1+r_+$. The contribution from
these terms makes the decay of the response function faster (relative to
$\alpha=0$) for small and large $\lambda$ and slower for intermediate
$\lambda$.
 
 The response function $\chi_{c}(t)$ is now calculated from
Eq.~\ref{eq:EQ43} using Eq.~\ref{eq:EQ9++++}, yielding to first order in
feedback
\begin{align}
 \chi_c (t>0)&  = \frac{\lambda}{(1+r_+)(1+r_+-\lambda)}\left( e^{-\lambda
 t}-e^{-(1+r_+)t}\right) \nonumber\\
 & + \alpha\left.\bigg( B_4e^{-(1+r_+)t}
 + C_4e^{-\lambda t} +
 D_4e^{-(1+r_++\lambda)t}  \right. \nonumber\\
 &
 \left. + E_4te^{-(1+r_+)t} + F_4te^{-\lambda t} \right.\bigg) + O(\alpha^2),
\label{eq:EQ43+++}
\end{align}
where $F_4<0$, while $E_4<0$ for $\lambda <1$ and $\lambda>1+r_+$ and $E_4>0$
for $1<\lambda<1+r_+$.

\subsection{Violation of the fluctuation-dissipation theorem}

 Having computed both the correlation and response functions for $S(t)$ and
$c(t)$ separately, we will now show that these quantities violate the
fluctuation-dissipation theorem (FDT) which holds for systems in thermal
equilibrium. In Fourier-space, the FDT reads (with the Boltzmann constant
$k_B$ and the temperature $T$)~\cite{Landau}
\begin{equation}
\bar{\omega}\tilde{{\cal
C}}(\bar{\omega})=2k_BT\mathrm{Im}\tilde{\chi}(\bar{\omega}),
\label{eq:FDT_general}
\end{equation}
if the correlation function $\tilde{{\cal C}}(\bar{\omega})$ and the response
function $\tilde{\chi}(\bar{\omega})$ refer to conjugate stimulus and response
variables.
 
 To relate to the system considered in this article, an energy
scale $\Delta U$ is introduced by writing the ratio of the transition rates
(with $c=0$) as $R_+/R_-^0\equiv r_+ = \exp(-\beta\Delta U)$, where $\beta$ is
the inverse temperature and $\Delta U$ is the energy difference between open
and closed channel states. Let us now introduce a 'field' $h$ which changes
the energy gap (analogous to an external magnetic field in the context of
magnetic models): $\Delta U\to\Delta U+h$, so that the dimensionless rate
$r_+$ is changed to $r_{+}^{\prime}=r_{+}e^{-\beta h}\approx r_{+}(1-\beta h)$
(where we have assumed that $h$ is small). The relation between the energy
change $h$ and the (dimensionless) change in the flipping rate $\phi$ (which
we defined earlier as the stimulus (Eq.~\ref{eq:EQ4b})) is therefore,
$\phi=-\beta r_{+}h$. Using this to transform our (dimensionless) response
function $\chi_S(t)$, Eq.~\ref{eq:FDT_general} can be written dimensionless as
\begin{equation}
\omega\tilde{\cal C}_S(\omega)=-2r_{+}\mathrm{Im}\tilde{\chi}_S(\omega),
\label{eq:FDT}
\end{equation}
and analogously for $c$.
 
 In the absence of feedback, we have
\begin{eqnarray}
\tilde{\cal C}_{S}(\omega) & = & \frac{2r_+}{(1+r_+)[(1+r_+)^2+\omega^2]},
\nonumber\\
 \tilde{\chi}_{S}(\omega) & = & \frac{1}{(1+r_+)(1+r_++i\omega)},
\label{eq:FDTx}
\end{eqnarray}
from which we find (unsurprisingly) that the FDT
 holds for $S(t)$. However,
for
 $c(t)$ the analogous results are (with $\alpha=0$)
\begin{eqnarray}
\tilde{\cal C}_{c}(\omega) & = &
\frac{2\lambda^2r_+}{(1+r_+)[(1+r_+)^2+\omega^2](\lambda^2+\omega^2)},
\nonumber \\
 \tilde{\chi}_c(\omega) & = &
\frac{\lambda}{(1+r_+)}\frac{1}{(\lambda+i\omega)(1+r_++i\omega)},
\label{eq:FDT++}
\end{eqnarray}
implying that FDT is violated for $c(t)$, i.e. the dynamics of $c$ cannot be
represented by an equilibrium system with the same temperature $T$ that
governs the dynamics of $S$.
 
 In order to characterize this violation
quantitatively, it is convenient to introduce an 'effective temperature'
$T^{\mathrm{eff}}$, defined as the ratio between the left and right hand sides
of Eq.~\ref{eq:FDT}, multiplied by $T$~\cite{JULICHER}:
\begin{equation}
\frac{T_{S}^{\mathrm{eff}}(\omega)}{T} = -\frac{\omega\tilde{{\cal
C}}_S(\omega)}{2r_+\mathrm{Im}\tilde{\chi}_S(\omega)}
\label{eq:T_eff}
\end{equation}
(analogously for $c$). $T$ is the 'real' temperature as introduced above. For
$\alpha=0$, from Eq.~\ref{eq:FDTx} and
 Eq.~\ref{eq:FDT++}, we then find that
$T_{S}^{\mathrm{eff}}/T=1$ for $S$ (i.e. FDT is fulfilled), while
$T_{c}^{\mathrm{eff}}/T=\lambda/(1+\lambda+r_+)<1$ for calcium (i.e. FDT is
violated). FDT is, however, asymptotically recovered in the limit $\lambda\gg
1+r_+$ (which is the limit where $c(t)$ follows $S(t)$ closely, see
Sec.~\ref{sec:averages}).
 
\begin{figure}[h!]
\psfrag{omega}[cc]{$\quad \omega$}
\psfrag{TeffS/T}{$T_{S}^{\mathrm{eff}}/T$}
\psfrag{Teffc/T}{$T_{c}^{\mathrm{eff}}/T$}
 \psfrag{a=.1}[cb]{$\scriptstyle
\alpha=0.1$}
 \psfrag{a=0}[cb]{$\scriptstyle \alpha=0$}
\includegraphics[width=.75\columnwidth]{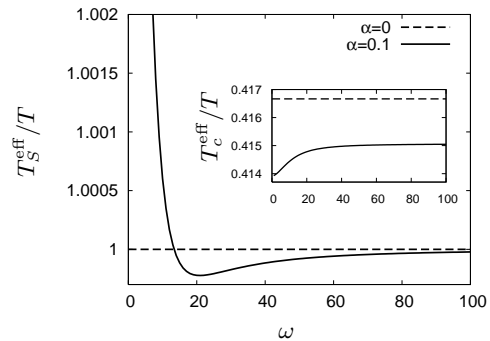}
\caption{The ratio $T^{\mathrm{eff}}/T$ (defined in Eq.~\ref{eq:T_eff}; with
the Fourier transforms of Eqs.~\ref{eq:EQ37},~\ref{eq:EQ37+},~\ref{eq:EQ43},
and~\ref{eq:EQ43+++}) of effective and 'real' temperature for the channel
state and calcium (inset) is plotted as a function of the frequency $\omega$
with and without feedback for $r_+=6$ and $\lambda=5$. Note that the
asymptotic limit in the case of calcium is different for the two cases and
differs from 1.}
\label{fig:fig5}
\end{figure}

 The ratio of effective-to-real temperature,
$T^{\mathrm{eff}}(\omega)/T$, is shown in Fig.~\ref{fig:fig5} as a function of
$\omega$ for both $S$ and $c$. We see that while the ratio for the channel
state approaches unity as $\omega\to\infty$, the corresponding curve for
calcium has a different asymptotic limit. The system as a whole is therefore
out of equilibrium already if no feedback is present. With feedback, at least
the $S$ variable becomes equilibrated in the limit of high frequencies.
 
%%%%%%%%%%%%%%%%%%%%%%%%%%%  Numerical results  %%%%%%%%%%%%%%%%%%%%%%%%%%%  

\section{Numerical results and Discussion}
\label{sec:numerics}

 We carried out numerical simulations in order to verify our analytical
results in the weak feedback limit, as well as to study the behavior of the
system in the intermediate ($\alpha \simeq 1$) and strong ($\alpha\gg 1$)
feedback cases. The dimensionless version of the system described by
Eqs.~\ref{eq:EQ5}--\ref{eq:EQ4b} was simulated using a fixed discrete time
step $\Delta t$~\footnote{The time step was chosen to be much smaller than all
the relevant timescales: $\Delta t\ll$ min$\{
\frac{1}{r_+},\frac{1}{1+\alpha},\frac{1}{\lambda} \}$. The kinetics of the
(dimensionless) calcium concentration is computed using a simple Euler forward
step algorithm. (Unless otherwise indicated, we used $\Delta t=10^{-3}$ and
the curves presented are averages over $10^6$ independent runs.) Feedback
strengths $\alpha=0, 0.1, 1, 10$ and $100$ were used, and $r_+$ values were
chosen from the range $[0.01,10000]$.}\footnote{In principle, a Gillespie-type
algorithm~\cite{Gillespie} could be used. Its implementation would, however,
require numerically inverting the dwelling time distribution
$P_1(t)=\exp\{-\frac{1}{\lambda}[\alpha (c(\tau_j)-1)](1-e^{-\lambda
t})-(1+\alpha)t\}$ for the open state for each value of $c$ (see
also~\cite{MAZZAG}). It is not clear, if this approach would decrease computer
time compared to the fixed time step algorithm.}.

 In general, the
feedback effects were found to increase in significance with larger
$\lambda$. However, very high values of $\lambda$ also tend to make the
kinetics of $c$ too closely tied to that of $S$ (see Fig.~\ref{fig:fig3}), and
for this reason, we performed most of our simulations with the intermediate
value $\lambda=5$ (unless otherwise indicated). The weak-feedback
($\alpha=0.1$) case was used for detailed comparison with the analytical
results. We observed from the data that the first order perturbation theory
works well up to $\alpha\approx 0.2$. Since the difference between the curves
for $\alpha=0$ and $\alpha=0.1$ is small, in the following, the $\alpha=0.1$
curve is not shown in the main figures. Instead, the comparison between
numerically obtained data points (symbols) and the analytical expressions
(lines) is shown in the insets of the figures as differences between the cases
with $\alpha=0.1$ and $\alpha=0$. The lines in the main figures connect
numerically obtained data points~\footnote{The data points in the main figures
(many hundreds in the plots over time, approximately 200 in the plots over
$r_+$) are equally spaced on the linear or logarithmic scale. Through
variation of $\Delta t$ and the number of independent runs used to calculate
the averages, it was checked, that the systematic error due to
time-discretization and restricted sampling size is in the range of line
thickness. In the insets as well as in the main figures of
Figs.~\ref{fig:fig15} and~\ref{fig:fig16}, this error is smaller than or
comparable to the symbol size.}.
 
\begin{figure}[h!]
\psfrag{r+}[cc]{$r_+$}
 \psfrag{<S>SS}{$\langle S\rangle$}
\psfrag{a=100}[cb]{$\scriptstyle \alpha=100$}
\psfrag{a=10}[cb]{$\scriptstyle \alpha=10$}
 \psfrag{a=1}[cb]{$\scriptstyle
\alpha=1$}
 \psfrag{a=0.1}[cb]{$\scriptstyle \alpha=0.1$}
\psfrag{a=0}[cb]{$\scriptstyle \alpha=0$}
\includegraphics[width=.75\columnwidth]{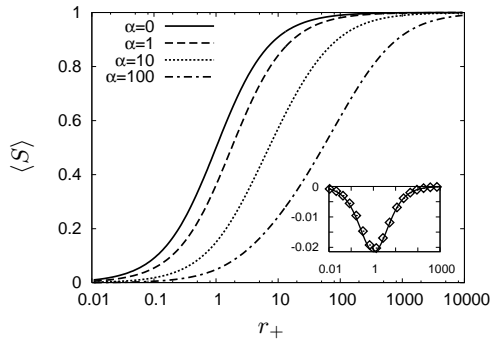}
\caption{The numerically obtained open channel fraction $\langle S\rangle$ in
the steady state plotted as a function of the opening rate
$r_+$. \textbf{Inset:} Difference between the cases $\alpha=0.1$ and
$\alpha=0$. The line represents the term linear in $\alpha$ from the
analytical expression in Eq.~\ref{eq:EQ35+}. For the high $r_+$ values as well
as for $\alpha=100$, a smaller time step of $\Delta t=10^{-4}$ was used.}
\label{fig:fig7}
\end{figure}

 In Fig.~\ref{fig:fig7}, the steady state open fraction $\langle S\rangle$
(which is the same as $\langle c\rangle$) is plotted as a function of the
opening rate $r_+$, for various values of $\alpha$. This is analogous to a
dose-response curve, if the open channel fraction (or, equivalently, the mean
Ca$^{2+}$) is interpreted as the response of the system to an external
stimulus that modifies the opening rate of the channels. We observe that
feedback shifts the response towards higher $r_+$ values by closing the
channels more often. However, the `dynamic range' of sensitivity (e.g. the
range of $r_+$ covered between 5\% and 95\% of the response) is found to be
increased with feedback: the curves become less steep for large $\alpha$.
 
\begin{figure}[h!]
\psfrag{r+}[cc]{$r_+$}
 \psfrag{C(0)}{$(\delta S)^2$}
\psfrag{a=100}[cb]{$\scriptstyle \alpha=100$}
\psfrag{a=10}[cb]{$\scriptstyle \alpha=10$}
 \psfrag{a=1}[cb]{$\scriptstyle
\alpha=1$}
 \psfrag{a=0.1}[cb]{$\scriptstyle \alpha=0.1$}
\psfrag{a=0}[cb]{$\scriptstyle \alpha=0$}
 A
\includegraphics[width=.75\columnwidth]{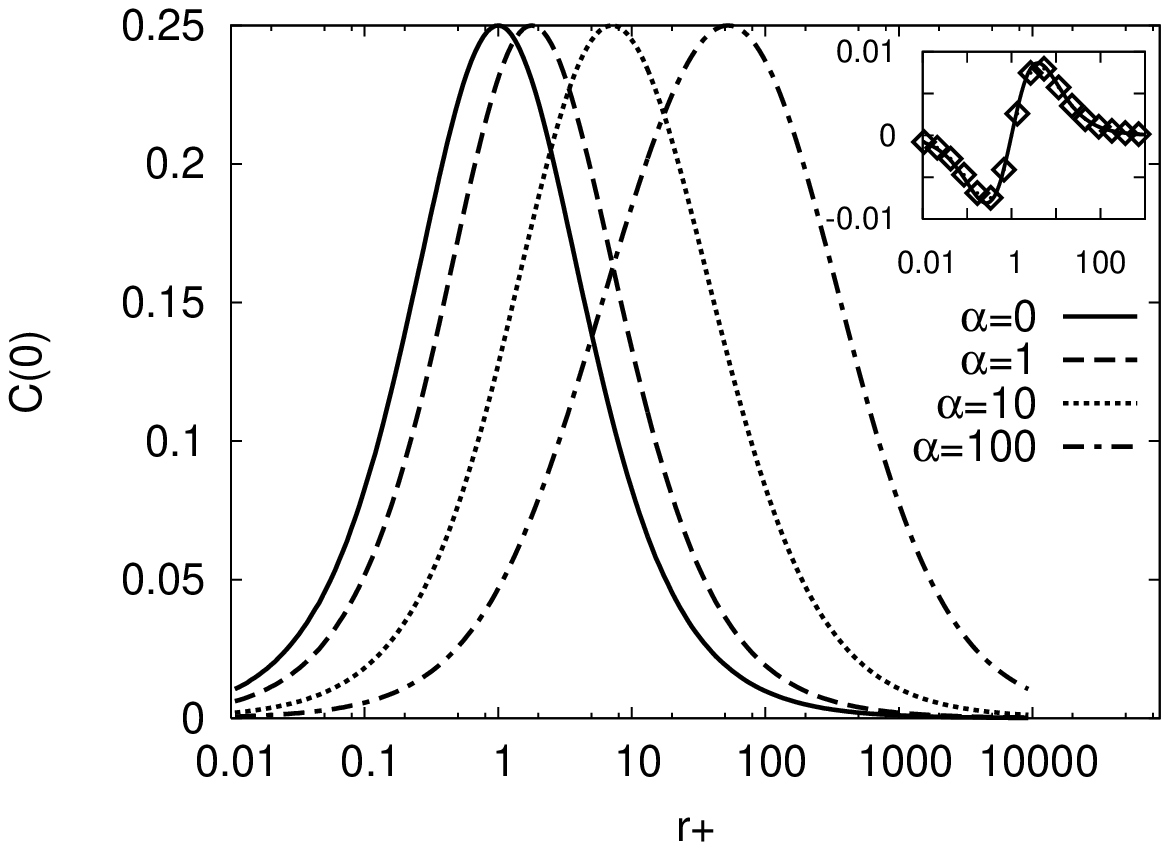}
 \psfrag{C*(0)}{$(\delta
c)^2$}
 
 B
 \includegraphics[width=.75\columnwidth]{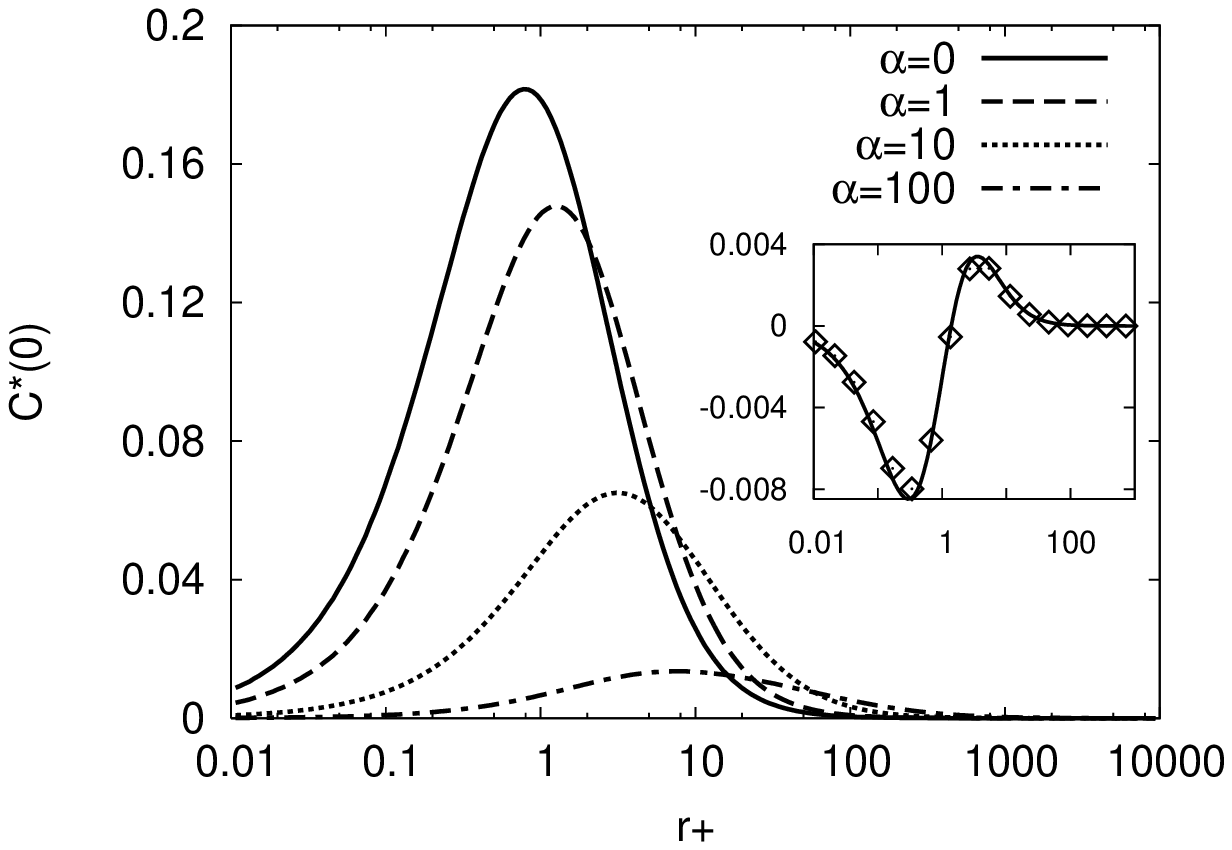}
\caption{Variance in $S$ (A) and $c$ (B) in the steady state plotted as
functions of $r_+$. \textbf{Inset:} Difference between the cases $\alpha=0.1$
and $\alpha=0$. The line represents the term linear in $\alpha$ from the
analytical expressions in Eqs.~\ref{eq:EQ37+++}. For the high $r_+$ values as
well as for the case $\alpha=100$ a smaller time step of $\Delta t=10^{-4}$
was used.}
\label{fig:fig12}
\end{figure}

In Fig.~\ref{fig:fig12} we show the variances $(\delta S)^2$ and $(\delta
c)^2$ of $S$ and $c$, respectively, in steady state as a function of $r_{+}$
for various feedback strengths. The fluctuations follow a bell-shaped curve as
a function of $r_+$, but the maximum shifts towards larger $r_+$ with
increasing feedback. As we remarked in Sec.~\ref{sec:averages}, the maximum of
the mean squared fluctuations $(\delta S)^2$ is always $\frac{1}{4}$
irrespective of the feedback strength (and occurs when $r_+$ is comparable to
the average closing rate $1 + \alpha \langle c \rangle$). For the $c$
fluctuations, however, we find that the fluctuations are generally suppressed
by feedback (except at very high $r_+$). The fluctuations again follow an
approximate bell-shaped curve, but the maximum sharply comes down with
increasing $\alpha$ and its position shifts to larger $r_+$. Note also that
for $\alpha >0$, the position of the peak for $c$-fluctuations always trails
the corresponding point for channel state. 
 
\begin{figure}[h!]
\psfrag{r+}[cc]{$r_+$}
 \psfrag{sqrt(C*(0))/<c>}{\quad $\delta c/\langle
c\rangle$}
 \psfrag{a=100}[cb]{$\scriptstyle \alpha=100$}
\psfrag{a=10}[cb]{$\scriptstyle \alpha=10$}
 \psfrag{a=1}[cb]{$\scriptstyle
\alpha=1$}
 \psfrag{a=0.1}[cb]{$\scriptstyle \alpha=0.1$}
\psfrag{a=0}[cb]{$\scriptstyle \alpha=0$}
\includegraphics[width=.75\columnwidth]{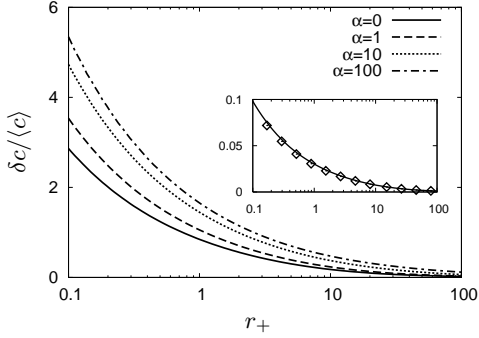}
\caption{The coefficient of variation plotted as a function of $r_+$ for
$\lambda=5$. \textbf{Inset:} Difference between the cases $\alpha=0.1$ and
$\alpha=0$. The line represents the term linear in $\alpha$ from the
analytical expression in Eq.~\ref{eq:EQ37+++-}. For the high $r_+$ values as
well as for the case $\alpha=100$ a smaller time step of $\Delta t=10^{-4}$
was used.}
\label{fig:fig13}
\end{figure}

\begin{figure}[h!]
\psfrag{r+}[cc]{$r_+$}
 \psfrag{sqrt(C*(0))/<c>}{\quad $\delta c/\langle
c\rangle$}
 \psfrag{a=100}[cb]{$\scriptstyle \alpha=100$}
\psfrag{a=10}[cb]{$\scriptstyle \alpha=10$}
 \psfrag{a=1}[cb]{$\scriptstyle
\alpha=1$}
 \psfrag{a=0.1}[cb]{$\scriptstyle \alpha=0.1$}
\psfrag{a=0}[cb]{$\scriptstyle \alpha=0$}
\includegraphics[width=.75\columnwidth]{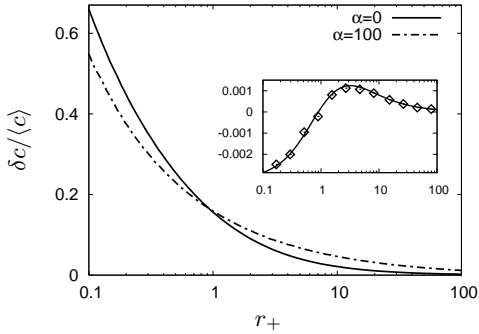}
\caption{The coefficient of variation as in Fig.~\ref{fig:fig13}, but with
$\lambda=0.05$ (only the curves for $\alpha=0$ and $100$ are shown for
clarity). Note that, in contrast to the previous case, the ratio is reduced
with increasing feedback for small $r_+$, which was also predicted from the
first order perturbative correction term in Eq.~\ref{eq:EQ37+++-}.}
\label{fig:fig14}
\end{figure}

 To investigate how fluctuations affect the precision of signal
transduction, we looked at the ratio of the standard deviation to the mean
(i.e. the noise-to-signal ratio in the steady state or the coefficient of
variation) for $c(t)$. In Fig.~\ref{fig:fig13} we show $\delta c/\langle
c\rangle$ plotted against $r_+$. For $\lambda=5$, we find that this ratio is a
monotonically decreasing function of $r_+$, independently of feedback. More
surprisingly, the ratio itself is enhanced by increasing feedback. Note that
for $r_+<0.1$, this ratio is of the order of 10 and is reduced to
''acceptable'' levels ($\delta c/\langle c\rangle\sim 1$) only for high
$r_+\sim 1-10$ depending on the feedback strength. However, the reverse effect
is observed when the pumping rate is reduced to $\lambda=0.05$. In this case,
for small $r_+$, the noise to signal ratio is reduced by feedback
(Fig.~\ref{fig:fig14}).
 
\begin{figure}[h!]
\psfrag{T}[cc]{$t$}
 \psfrag{C(T)}{${\cal C}_{S}(t)$}
\psfrag{a=10}[cb]{$\scriptstyle \alpha=10$}
 \psfrag{a=1}[cb]{$\scriptstyle
\alpha=1$}
 \psfrag{a=0.1}[cb]{$\scriptstyle \alpha=0.1$}
\psfrag{a=0}[cb]{$\scriptstyle \alpha=0$}
 A
\includegraphics[width=.75\columnwidth]{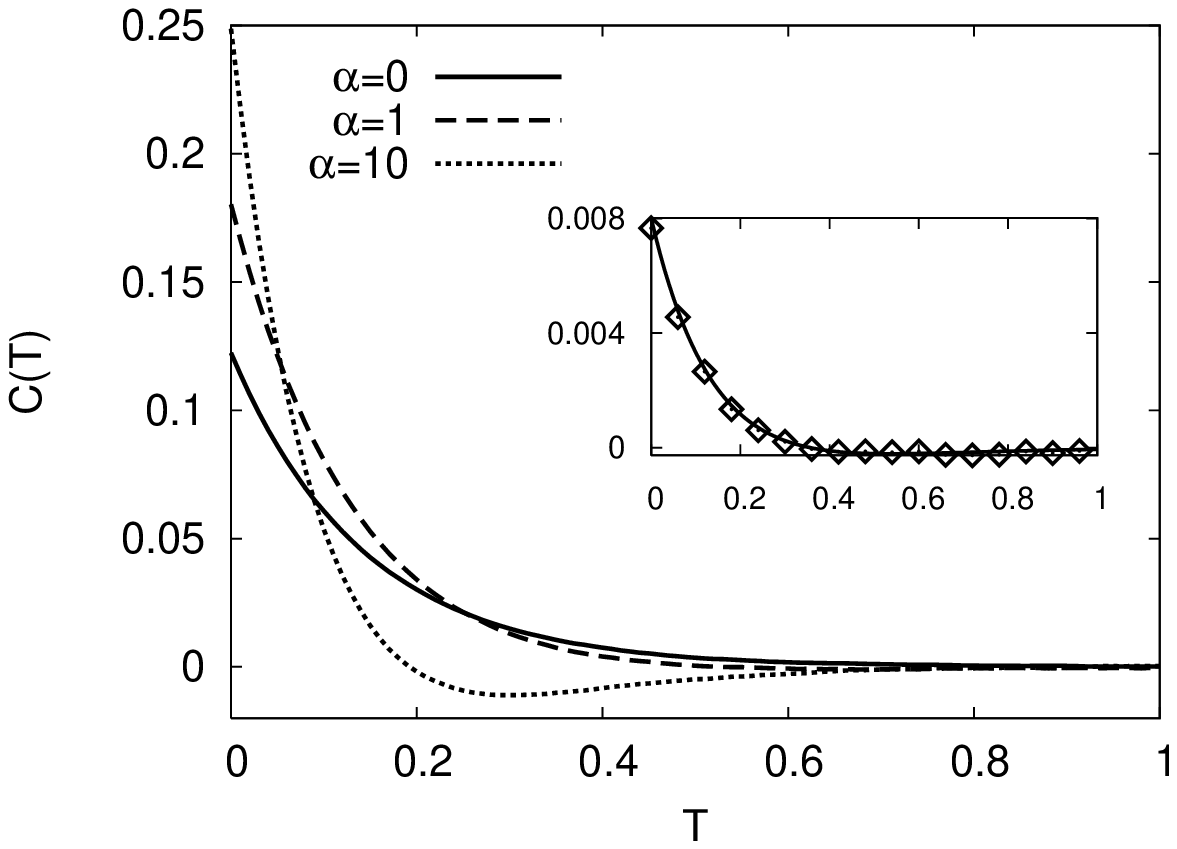}
 \psfrag{C*(T)}{${\cal
C}_{c}(t)$}
 
 B
 \includegraphics[width=.75\columnwidth]{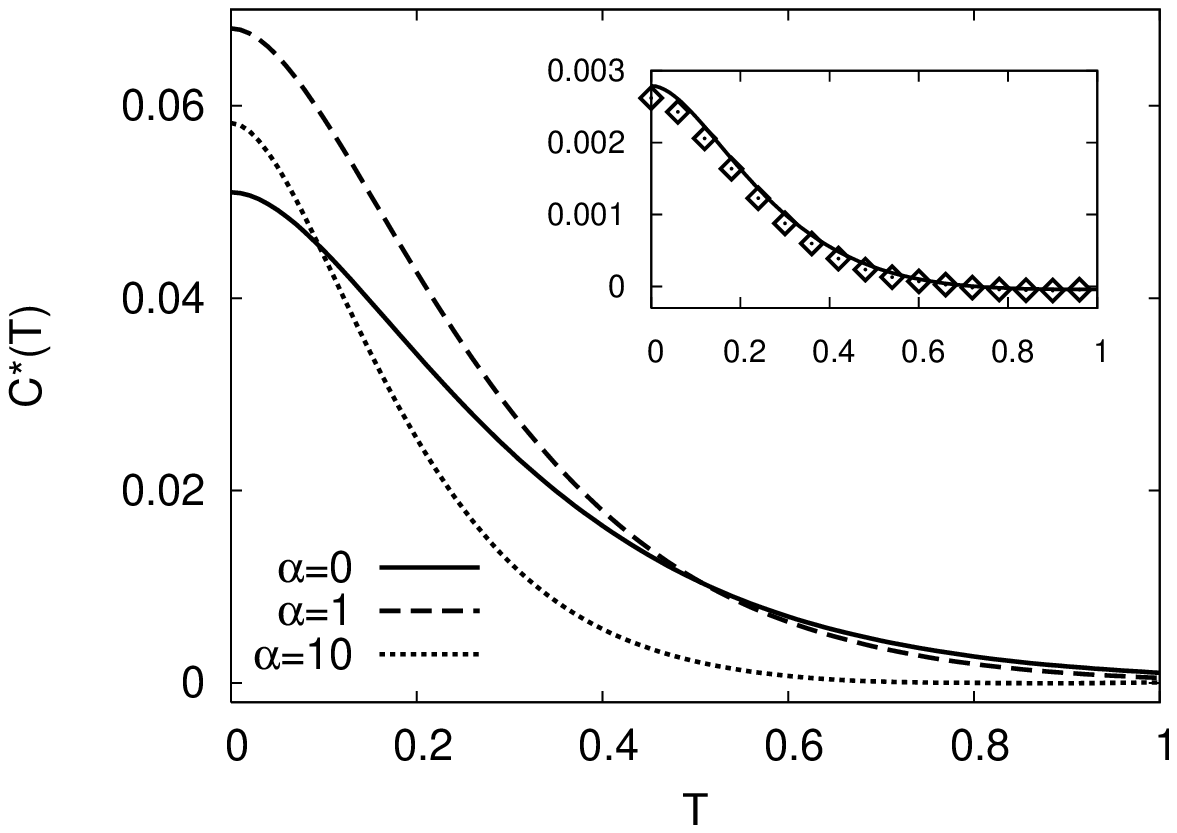}
\caption{Auto-correlation function for $S(t)$ (A) and $c(t)$ (B) in the steady
state for $\lambda=5$ and $r_+=6$. \textbf{Inset:} Difference between the
cases $\alpha=0.1$ and $\alpha=0$. The line represents the term linear in
$\alpha$ from the analytical expressions for ${\cal C}_{S}(t)$
(Eq.~\ref{eq:EQ37}) and  ${\cal C}_{c}(t)$ (Eq.~\ref{eq:EQ37+}).}
\label{fig:fig9}
\end{figure}

 We now turn to time-dependent quantities. Figs.~\ref{fig:fig9} A and B show
the auto-correlation functions for the channel state and calcium,
respectively, in the case of strong signal ($r_+=6$). In general, one notes
that the effect of increasing feedback is to reduce the characteristic time
scale of decay of the correlations. 
 When the feedback parameter is very
large, the channel auto-correlation function briefly becomes negative before
vanishing at longer times. This anti-correlation in the channel state may be
interpreted physically as the rapid closing of an open channel by the
Ca$^{2+}$ which enters through it. In other words, in the presence of
feedback, a channel which is open at any point of time has a reduced open
probability at subsequent times (compared to the unconditioned mean open
probability). By contrast, the auto-correlation function for $c(t)$ for the
same parameters decays monotonically with time. 
 In the cases of weak signal
($r_+=0.5$) or high pump rate ($\lambda=50$), both the channel state and
calcium auto-correlation functions decay monotonically (data not shown). 
 
\begin{figure}[h!]
\psfrag{omega}[cc]{$\quad \omega$}
 \psfrag{Pc}{$P_c$}
\psfrag{PS}[cc]{$P_{S}$}
 
 \psfrag{a=10}[cb]{$\scriptstyle \alpha=10$}
\psfrag{a=0.2}[cb]{$\scriptstyle \alpha=0.2$}
 \psfrag{a=0}[cb]{$\scriptstyle
\alpha=0$}
 \psfrag{1e-04}[cc]{$\scriptstyle 10^{-4}$}
\psfrag{o-4}{$\omega^{-4}$}
 \psfrag{o-2}[cc]{$\omega^{-2}$}
\includegraphics[width=.75\columnwidth]{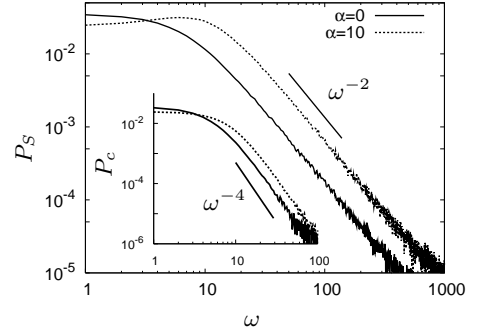}
\caption{Power spectral densities for the channel state (main figure) and
calcium concentration (inset) are plotted against the angular frequency
$\omega$ with and without feedback on a logarithmic scale. The parameter
values used are $r_+=6$ and $\lambda=5$. The straight lines are guides to the
eye. To get results also for high frequencies, a small time step of $\Delta
t=10^{-5}$ was used.}
\label{fig:fig10}
\end{figure}

In Fig.~\ref{fig:fig10}, we plot the power spectral density for calcium and
the channel state calculated from the auto-correlation function using
Eq.~\ref{eq:EQ9+-}, in the presence and absence of feedback. We observe that
irrespective of feedback, $P_{c}(\omega)\sim \omega^{-4}$ and
$P_{S}(\omega)\sim \omega^{-2}$ as $\omega\to\infty$. It may also be noted
that $P_{S}(\omega)$ is non-monotonic in this plot, and the observed peak is
consistent with the dip in the corresponding auto-correlation function in $S$
(Fig.~\ref{fig:fig9}A). Notice also that a similar peak is absent for $P_{c}$,
which is also consistent with the fact that ${\cal C}_{c}(t)$ is monotonic
(Fig.~\ref{fig:fig9}B). 
In a different context, the reduction in noise strength and the shifting of
the peak of the power-spectrum from low to high frequencies with increasing
feedback has also been observed in simulations of a model of an autoregulated gene circuit~\cite{Simpson:2003} with negative feedback. 
An instructive discussion of the the general conditions under which feedback can lead to a reduced noise intensity and increased noise bandwidth is given in the Appendix of Ref.~\cite{Simpson:2003}. 

\begin{figure}[h!]
\psfrag{T}[cc]{$t$}
 \psfrag{G01(0,T;0)}{\hspace{-.5cm}$G_{01}(0,c_0=0;t)$}
\psfrag{a=10}[cb]{$\scriptstyle \alpha=10$}
 \psfrag{a=1}[cb]{$\scriptstyle
\alpha=1$}
 \psfrag{a=0.1}[cb]{$\scriptstyle \alpha=0.1$}
\psfrag{a=0}[cb]{$\scriptstyle \alpha=0$}
 A 
\includegraphics[width=.75\columnwidth]{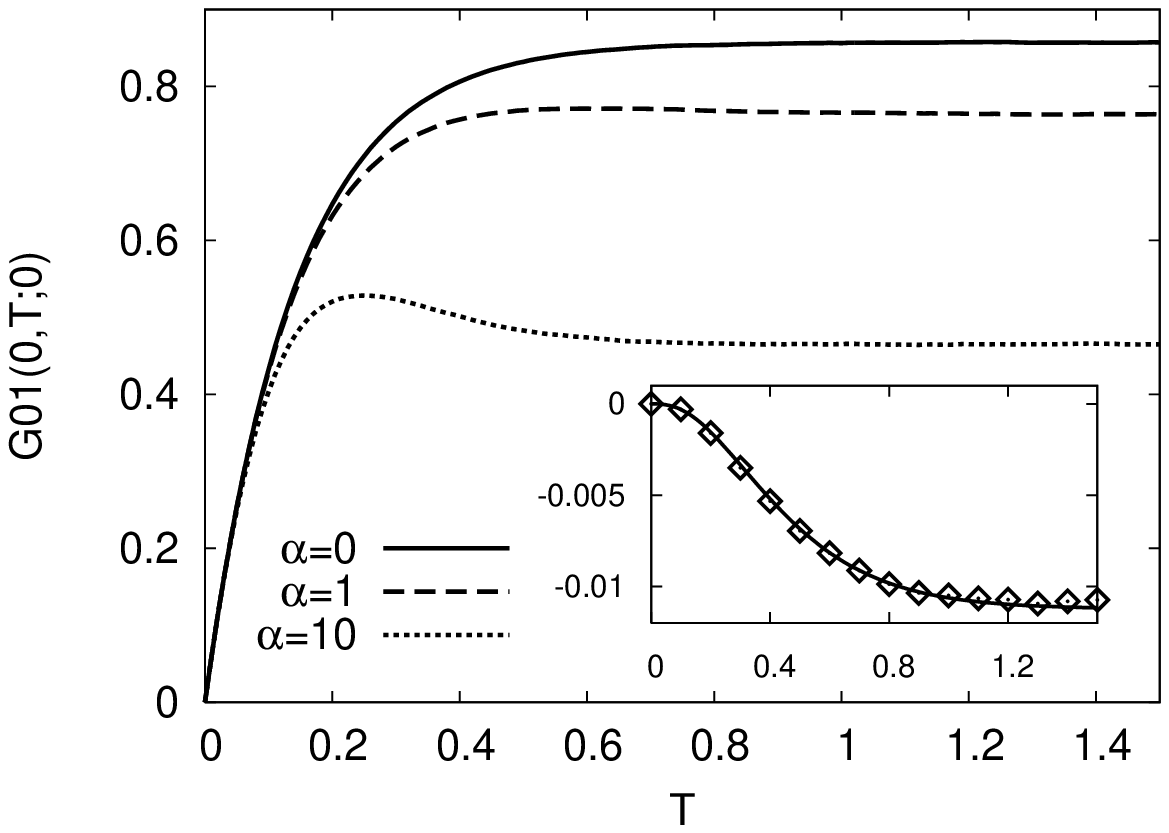}
\psfrag{Gc01(0,T;0)}{$\langle c(t)\rangle$}
 
 B
\includegraphics[width=.75\columnwidth]{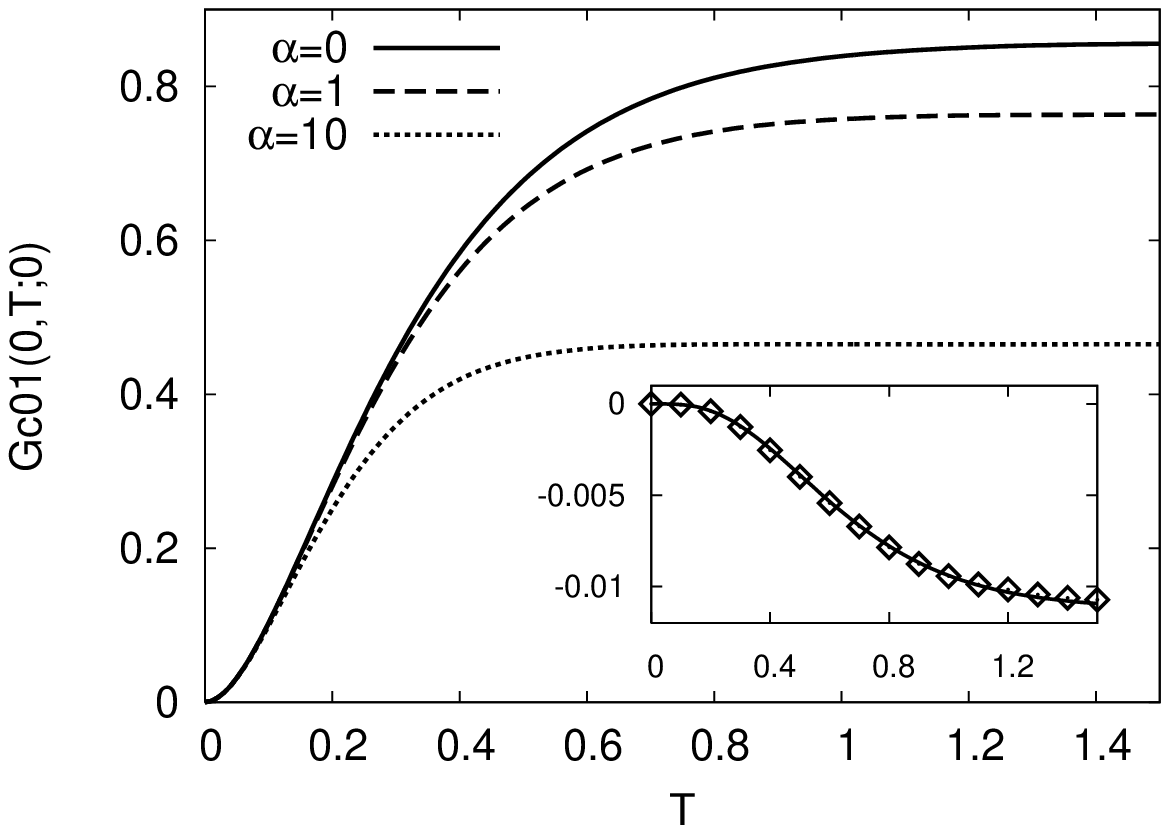}
\caption{Time evolution of the open-probability (A) and the mean calcium
concentration (B) when the channel starts in the closed state
($S(t=0)=c(t=0)=0$) for different values of the feedback parameter
$\alpha$. The opening rate is $r_+=6$ and $\lambda=5$. \textbf{Inset:}
Difference between the cases $\alpha=0.1$ and $\alpha=0$. The solid line
represents the term linear in $\alpha$ from the analytical expressions
Eq.~\ref{eq:EQ26} ($G_{01}^{(1)}(0,0,t)$) and Eq.~\ref{eq:EQ8}.}
\label{fig:fig6}
\end{figure}

\begin{figure}[h!]
\psfrag{T}[cc]{$t$}
 \psfrag{chi(T)}{$\chi_{S}(t)$}
\psfrag{a=10}[cb]{$\scriptstyle \alpha=10$}
 \psfrag{a=1}[cb]{$\scriptstyle
\alpha=1$}
 \psfrag{a=0.1}[cb]{$\scriptstyle \alpha=0.1$}
\psfrag{a=0}[cb]{$\scriptstyle \alpha=0$}
\includegraphics[width=.75\columnwidth]{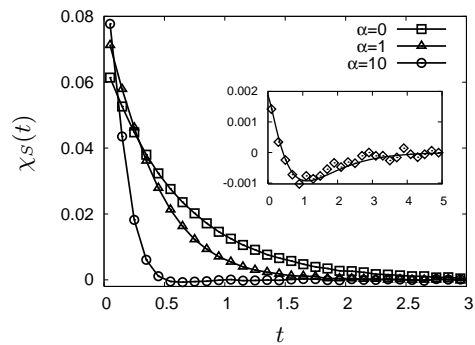}
\caption{The linear response function of the system, computed numerically
using a step stimulus (see text) and a numerical differentiation for various
values of feedback ($\lambda=5$). The points shown are the averages over
$10^9$ independent runs (symbols are used to indicate the larger error in this
calculation -- the lines in the main figure are guides to the
eye). \textbf{Inset:} The difference between $\alpha=0.1$ and $\alpha=0$ is
compared with the analytical expression in Eq.~\ref{eq:EQ43} (line).}
\label{fig:fig15}
\end{figure}

\begin{figure}[h!]
\psfrag{omega}[cc]{$\quad \omega$}
 \psfrag{A}{$A$}
\psfrag{a=10}[cb]{$\scriptstyle \alpha=10$}
 \psfrag{a=1}[cb]{$\scriptstyle
\alpha=1$}
 \psfrag{a=0.1}[cb]{$\scriptstyle \alpha=0.1$}
\psfrag{a=0}[cb]{$\scriptstyle \alpha=0$}
 A
\includegraphics[width=.75\columnwidth]{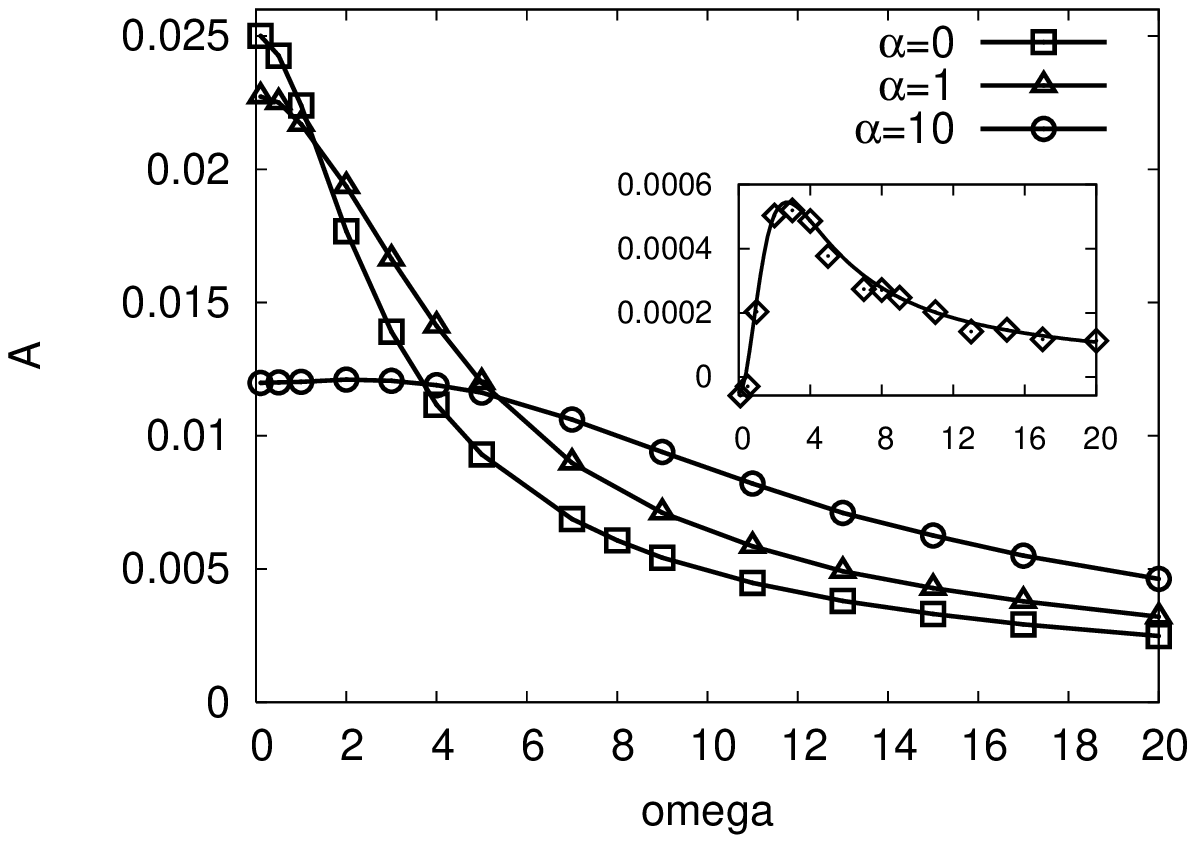}
\psfrag{theta/pi}{$\theta/\pi$}
 
 B
\includegraphics[width=.75\columnwidth]{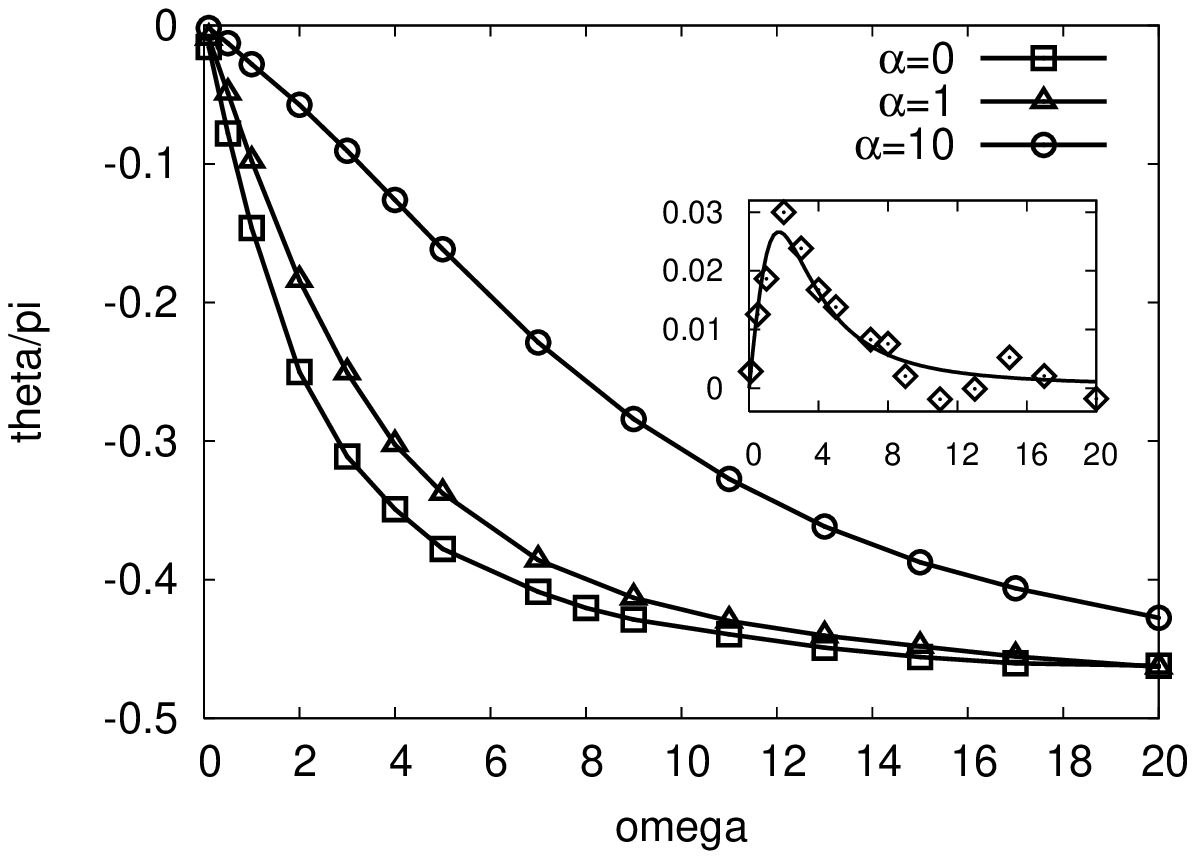}
\caption{Amplitude (A) and phase (B) of the response $\langle
S(t)\rangle^{\phi}$ to a sinusoidal signal $\phi(t)=a\sin(\omega t)$ with
amplitude $a=0.1$ and baseline $r_+=1$, plotted as a function of stimulus
angular frequency $\omega$ ($\lambda=5$). The lines in the main figures are
guides to the eye. \textbf{Inset:} Difference between the cases $\alpha=0.1$
and $\alpha=0$.  The line represents the $O(\alpha)$ correction computed
analytically (not shown in text). Note the regime of linear decay of $\theta$
versus $\omega$ in B for small $\omega$, which signifies a constant time delay
in the response.}
\label{fig:fig16}
\end{figure}

 Fig.~\ref{fig:fig6} shows the time evolution of the open-probability of the
channel (Fig.~\ref{fig:fig6}A) and the mean calcium concentration
(Fig.~\ref{fig:fig6}B) for different values of the feedback parameter $\alpha$
when the channel was initially closed and $c(t=0)=0$. Note that the relaxation
to the steady state is monotonic for small $\alpha$, but for higher $\alpha$,
the response in the channel (Fig.~\ref{fig:fig6}A) exhibits a maximum, and
then settles into the steady state, while the response of $c$ does not show
this behavior (Fig.~\ref{fig:fig6}B). 
 
 In Fig.~\ref{fig:fig15}, we show the linear response function
$\chi_{S}(T)$ for different values of $\alpha$. In order to compute the
response function numerically, the system was evolved with $r_+=0.5$ until it
reached the steady state. Then, $r_+$ was increased to $0.6$ and the response
function was calculated using Eq.~\ref{eq:EQ9+++}. By definition, the response
function shows how a sharply peaked input stimulus ($\phi(t)\sim \delta(t)$)
is `transmitted' across the channel. We observe that increasing the feedback
reduces the time constant of decay, making the output signal sharper and more
similar to the input, when $\lambda$ is relatively high. This is in agreement
with our assertion that high values of $\lambda$ improve the short-time scale
response characteristics of the system. We also confirmed this by studying the
effect of $\lambda$ on the response time when feedback is strong: in this
case, reducing $\lambda$ was found to increase the response time appreciably 
(for example, when $\alpha=10$, a 50-fold reduction in $\lambda$ was observed to almost
double the response time).

 We used the channel linear response function derived in
Sec.~\ref{sec:responsefunctions} to compute the response of the system to a
periodic stimulus of the form $\phi(t)= a\sin \omega t$ for 14 different
values of $\omega$ (Fig.~\ref{fig:fig16}). An explicit calculation shows that
the linear channel response has the form $\langle S(t)\rangle^{\phi}-\langle
S\rangle=A\sin(\omega t+\theta)$, where both the amplitude $A$ and the phase
lag $\theta$ are frequency-dependent: $A(\omega)
=a/(1+r_+)[(1+r_+)^2+\omega^2]^{-1/2}+O(\alpha)$ and
$\theta(\omega)=-\frac{\pi}{2}+\arctan[(1+r_+)/\omega]+O(\alpha)$, omitting
the $O(\alpha)$ terms for the sake of brevity. The complete expressions
(including the first order corrections (to be found at~\cite{Borowski})) for $A$ and $\theta$ are checked against numerical simulations in the insets of Figs.~\ref{fig:fig16} A and B. The amplitude is always a decreasing function of the frequency, however, the feedback has the effect of increasing the amplitude of the response relative to the no-feedback case at sufficiently high frequencies. At lower frequencies, the opposite effect is observed: introducing feedback tends to reduce the amplitude of the output, however, the decay as a function of the external frequency is more gradual. In simple terms, the over-all effect of the feedback here is to widen the range of frequencies effectively sensed by the system. The increase in response at large frequencies is consistent with the reduction in the time constant of the linear response function which we noted earlier. At small frequencies, on the other hand, the overall adaptation produced by the negative feedback leads to a drop in the response.

The phase shift of the response decreases monotonically from 0 to $-\frac{\pi}{2}$ as $\omega$ is increased, although the change is more gradual when feedback is present. Note that for large $\alpha$, there is a significant linear regime in the $\theta-\omega$ curve, which means that in this regime, the time delay of the response is effectively independent of the stimulus frequency.

 To summarize, numerical simulations provide excellent support for the
analytical predictions from the path-integral theory in the weak feedback
limit. In addition, we observed qualitatively new features in the
auto-correlation function and in the relaxation of the mean open probability
to the steady state when feedback is strong. In particular,
Figs.~\ref{fig:fig13},~\ref{fig:fig14},~\ref{fig:fig9} and~\ref{fig:fig16} are
the principal results of this paper, and will be discussed more in the next
section in the context of experimental results.

%%%%%%%%%%%%%%%%%%%%%%%%%%%  Conclusions  %%%%%%%%%%%%%%%%%%%%%%%%%%%  

\section{Conclusions}
\label{sec:conclusions}

 In this paper, we studied the stochastic kinetics of a simple
auto-regulatory signaling module with negative feedback using path-integral
techniques and numerical simulations. Within our formalism, all the
statistical averages and correlation functions that characterize the long-time
kinetics of the system are formally expressed as a power series in the
feedback parameter. Explicit expressions were obtained for mean values and
correlation functions to first order in the feedback parameter. These
analytical results were compared to numerical simulations. The simulations
were also done in the strong feedback regime beyond the applicability of
perturbation theory. In particular, we investigated how the system responds to
temporal variations in an external stimulus, and how the feedback regulates
this response.
 
 Our principal conclusions from this study may be outlined
as follows. We find that, in the context of the ion-channel model, the rate of
removal of Ca$^{2+}$ ions, $\lambda$, is the single most important control parameter
of the model. When $\lambda$ is large (compared to the intrinsic closing rate for the channel), 
the short-time scale dynamic response
characteristics of the system are improved by negative feedback. 
In this case, feedback reduces the time scale of decay of the auto-correlation and
response functions, and thus extends the response of the system to higher
frequencies of input stimulus. 

For all parameter values, the negative feedback shifts the range of sensitivity of the system (i.e. the range of stimulus magnitudes for which the response is appreciably non-zero and not saturated) towards stronger stimuli. At the same time, the range of sensitivity gets broader (i.e., the "dynamic range" is increased) with increasing feedback. While the steady state response to a fixed stimulus is always reduced by the negative feedback, a remarkable effect was seen in the 
coefficient of variation for the calcium concentration. Increasing the feedback was found to increase this
quantity in the large $\lambda$ regime. For small
$\lambda$, however, the opposite was seen: the coefficient of variation was
reduced with feedback, for sufficiently weak stimuli. In this regime, the negative feedback therefore improves 
the precision with which one can discriminate among stimuli of different magnitudes based on the system output. 
Having a small $\lambda$, however, was generally found to adversely affect the temporal response characteristics.

Starting from the experimental literature, we have obtained estimates for the dimensionless parameters in our model for the specific case of 
calcium signaling in olfactory sensory neurons (the corresponding module is shown in the lower part of Fig.~\ref{fig:fig1}). 
The rate of opening of the cyclic-nucleotide-gated channel varies 
in the range $r_+ \approx 10^{-5}$
-- $20$ depending on the strength of the odorant stimulus. The parameter governing the
calcium dynamics is $\lambda\approx 20$ and for the feedback parameter we find
$\alpha\approx 1$ -- $10$~\footnote{
To obtain these estimates, we view the olfactory cilium as consisting of a series of cylindrical segments, each containing one cyclic-nucleotide-gated  (CNG) channel. 
On the time scales of interest, we neglect calcium diffusion  between neighboring segments. We estimate the parameter values in each segment as follows: $J\approx 1.3\cdot
10^{-19}\mathrm{mol \cdot s^{-1}}$~\cite{Reisert2003}; $V\approx 6\cdot
10^{-24}\mathrm{m^3}$ (calculated from the density of CNG channels and the
geometry of the cilium as cited in~\cite{REIDL2005}, assuming a homogeneous
distribution of channels); $\bar{\lambda}\approx 10^4\mathrm{s^{-1}}$
(calculated from a Taylor expansion of the calcium-extrusion term used
in the deterministic model of Ref.~\cite{REIDL2005});
$R_-^0\approx 500\mathrm{s^{-1}}$ (from single channel measurements with
weak stimulus and low calcium (e.g.~\cite{Ruiz1999,Gavazzo2000})); $R_+\approx
5\cdot 10^{-3}$ -- $10^4\mathrm{s^{-1}}$ (based on an open probability of
$10^{-5}$ if no stimulus is present to $0.95$ for full
activation~\cite{Kaupp_Seifert_PhysiolRev_2002}); $\bar{\alpha}\approx 3\cdot
10^8$ -- $2\cdot 10^9\mathrm{s^{-1}M^{-1}}$ (based on data in the figures
of Refs.~\cite{Kleene1999}(lower estimate) and \cite{Bradley2004} (upper estimate)).}. 
Note, in particular,
that $\lambda$ is found to have a relatively large value. 
In the view of our results, this suggests that the calcium-mediated feedback in olfactory transduction evolved so as to improve the temporal response characteristics of the system, 
rather than to improve the discriminability of weak stimuli.

Our findings are consistent with the results of experimental studies in a variety of sensory systems with calcium-mediated negative feedback. 
The feedback-induced shift of the range of sensitivity to higher stimulus magnitudes has been demonstrated in olfactory sensory neurons and in cone photoreceptor cells (reviewed in Ref.~\cite{Bradley2005,Pifferi2006}). 
The temporal characteristic of response was found to be altered in several studies in which the strength of feedback was experimentally manipulated. 
In Ref.~\cite{Bhandawat}, the time course of response to a pulse stimulus was found to be slowed down
when the frog olfactory cilium was put in lowered external calcium concentrations (in such conditions, the ion flux J, and therefore the dimensionless coupling $\alpha$, is decreased). 
A more pronounced effect of this type was observed in similar experiments for Drosophila photoreceptor cells in Ref.~\cite{Henderson}. A marked slowing down of the response to a step stimulus was observed for the newt olfactory cilium 
in Ref.~\cite{Takeuchi2002} when the calcium flux J was decreased by changing the holding potential. Fluctuations in the steady state have received less attention in the biological literature. The power spectrum of odorant-induced current fluctuations 
across the membrane of rat olfactory sensory cilia was studied experimentally by Lowe and Gold~\cite{LOWE}. They observed that the tail of the power-spectrum decays with an effective exponent in the range 2.3 -- 2.5, which is suggestive 
of a crossover between $\omega^{-2}$ and $\omega^{-4}$ decay. 
In this case, however, the dynamics of the cAMP module (shown in the upper part of Fig.~\ref{fig:fig1}) should be taken into account in the theoretical treatment, 
as fluctuations in cAMP give an important contribution to the variability of the response \cite{LOWE}. 
 
 We are currently in the process
of extending this study to include spatial effects. In particular, in
olfactory sensory neurons, the Ca$^{2+}$ ion channels are spatially
distributed along long and thin cellular compartments called cilia. The
feedback studied in the present manuscript coupled with calcium diffusion
inside the cilium can give rise to non-trivial spatial correlations between
the channels, and it will be interesting to study how the spatial coupling
affects the overall kinetics of the system.
 
\begin{acknowledgments}
We gratefully acknowledge Peter H\"anggi and Benjamin Lindner for helpful and
illuminating discussions. We also thank an anonymous referee for bringing 
Ref.~\cite{IEEE} and the associated engineering literature to our attention.

\end{acknowledgments}

%%%%%%%%%%%%%%%%%%%%%%%%%%%  appendix  %%%%%%%%%%%%%%%%%%%%%%%%%%% 

\appendix

\section{Calculation of integrals $\boldsymbol{I_0}$, $\boldsymbol{I_1}$ and $\boldsymbol{I_2}$}
\label{sec:appendix_I}

 {\it Generalized convolution theorem}:
 
 The following general result is
very useful in performing many calculations using the path-integral
technique. If $g(t)=\int_{0}^{t}
f_1(t_1)\int_{0}^{t-t_1}f_2(t_2)\cdots\int_{0}^{t-\cdots -t_{n-1}}f_n(t_n)$,
then its Laplace transform $\tilde{g}(s)=\int_{0}^{\infty}g(t)e^{-st}dt$ is
given by
\begin{equation}
\tilde{g}(s)=\frac{1}{s}\prod_{i=1}^{n}\tilde{f}_{i}(s).
\label{eq:EQB0-}
\end{equation}
This result is proved in a straightforward way by repeated application of the
standard convolution theorem of Laplace transforms.
 
 {\it Calculation of
$I_0$}:
 
 The integral $I_0$ is relatively easy to compute. The Laplace
transform ${\tilde I}_0(s)$ can be written easily from the definition of the
integral in Eq.~\ref{eq:EQ21} and using the above theorem:
\begin{equation}
{\tilde I}_{0}(s;m)=s^{-(m+1)}(s+1-r_+)^{-m}.
\label{eq:EQB0}
\end{equation}

 {\it Calculation of $I_1$}:
 
 In order to calculate the integrals $I_1$
and $I_2$, it is necessary  to express $c(t)$ in terms of the time interval
variables using Eq.~\ref{eq:EQ8}.  The explicit relations are,
\begin{align}
c(\tau)= & 1-e^{-\lambda(\tau-\tau_1)}\quad \text{for}\quad \tau_1\leq
\tau\leq \tau_2, \nonumber\\
 c(\tau)= & 1+e^{-\lambda\tau}
\left(e^{\lambda\tau_2}-e^{\lambda\tau_1}+\cdots
+e^{\lambda\tau_{j-1}}-e^{\lambda\tau_j}\right)\nonumber \\
 & \quad
\mathrm{for} \quad \tau_j\leq \tau\leq \tau_{j+1}\quad \text{with odd}\quad
j\geq 3.
\label{eq:EQB1}
\end{align}

 From that it follows
\begin{equation}
c(\tau_i)=\sideset{}{''}\sum_{j=1}^{i-1}\left(
e^{-\lambda(\tau_i-\tau_{j+1})}-e^{-\lambda(\tau_i-\tau_{j})}
\right)~~\text{for even}~i\geq 2.~~~
\label{eq:EQB3}
\end{equation}
Using Eq.~\ref{eq:EQB3} in Eq.~\ref{eq:EQ21}, we write $I_1=g_1-g_2$, with
\begin{align}
 g_1(0,t;m)= & \sideset{}{'}\sum_{i=2}^{2m}\sideset{}{''}\sum_{j=1}^{i-1}\int
 {\cal D}T\times \nonumber\\
 & \exp\left(-\left[(1-r_+)\sum_{k=1}^{m}
 t^{\prime}_{k}+\lambda(\tau_i-\tau_{j+1})\right]\right)
\label{eq:EQB4}
\end{align}
and
\begin{align}
 g_2(0,t;m)= & \sideset{}{'}\sum_{i=2}^{2m}\sideset{}{''}\sum_{j=1}^{i-1}\int
 {\cal D}T\times \nonumber\\
 & \exp\left(-\left[(1-r_+)\sum_{k=1}^{m}
 t^{\prime}_{k}+\lambda(\tau_i-\tau_{j})\right]\right).
\label{eq:EQB5}
\end{align}

 Note that when $j\leq i-3$,$\tau_i-\tau_{j+1}=\sum_{(j+3)/2}^{i/2}\left(
t^{\prime}_{k}+t_{k}\right)$ and is zero when $j=i-1$. Let us therefore split
the second sum, and treat  these cases separately. In order to compute the
Laplace transforms, we use the theorem in Eq.~\ref{eq:EQB0-}. It is now
straightforward to write the Laplace transform of $g_1$ in Eq.~\ref{eq:EQB4}
using this theorem. The result is
\begin{align}
{\tilde g}_1 & (s)= ms^{-(m+1)}(s+1-r_+)^{-m} \nonumber\\ 
 & +
s^{-(m+1)}(s+1-r_+)^{-m}\sideset{}{'}\sum_{i=4}^{2m}y^{i/2}\sideset{}{''}\sum_{j=1}^{i-3}y^{-(j+1)/2},
\label{eq:EQB6}
\end{align}
where we have defined
$y=s(s+\lambda)^{-1}(s+1-r_+)(s+1-r_++\lambda)^{-1}<1$.
 The double geometric
sum in Eq.~\ref{eq:EQB6} turns out to be
$y(1-y)^{-1}\left[m+(1-y)^{-1}(y^m-1)\right]$, which gives
\begin{align}
{\tilde g}_1(s)= s^{-(m+1)} & (s+1-r_+)^{-m}\frac{1}{1-y}\times \nonumber \\
 & \left(m+\frac{y}{1-y}(y^m-1)\right).
\label{eq:EQB9}
\end{align}

 Calculation of $g_2$ also proceeds along similar lines. After some
straightforward algebra, we find that
\begin{eqnarray}
{\tilde
g}_2(s)=s^{-(m+1)}(s+1-r_+)^{-m}\frac{s+1-r_+}{s+1-r_{+}+\lambda}\times\nonumber
\\
 \left(
m+\sideset{}{'}\sum_{i=2}^{2m}y^{i/2}\sum_{l=1}^{i/2-1}y^{-l}\right).~~
\label{eq:EQB11}
\end{eqnarray}
The double sum is easily shown to be equal to
$y(1-y)^{-1}[m+(1-y)^{-1}(y^m-1)]$, and this gives
\begin{eqnarray}
{\tilde
g}_2(s)=s^{-(m+1)}(s+1-r_+)^{-m}\frac{s+1-r_+}{s+1-r_{+}+\lambda}\times\nonumber
\\
 \frac{1}{1-y}\left( m+\frac{y}{1-y}(y^m-1)\right).~~~
\label{eq:EQB12}
\end{eqnarray}
After putting together Eq.~\ref{eq:EQB9} and Eq.~\ref{eq:EQB12}, we arrive at
\begin{eqnarray}
{\tilde
I}_{1}(s;m)=s^{-(m+1)}(s+1-r_+)^{-m}\frac{s+\lambda}{2s+1-r_{+}+\lambda}\nonumber
\\
 \left( m+\frac{y}{1-y}(y^m-1)\right).~~~
\label{eq:EQB14}
\end{eqnarray}

 {\it Calculation of $I_2$}:
 
 For $I_2$, the following integral is
needed (obtained using Eqs.~\ref{eq:EQB1}):
\begin{eqnarray}
\int_{\tau_j}^{\tau_{j+1}}c(\tau)d\tau=t^{\prime}_{(j+1)/2}+
\frac{1}\lambda\left( e^{-\lambda\tau_j}-e^{-\lambda\tau_{j+1}}\right)
\times\nonumber\\
\left(\sideset{}{'}\sum_{k=2}^{j-1}e^{\lambda\tau_k}-\sideset{}{''}\sum_{l=1}^{j}e^{\lambda\tau_l}\right)~~~
\label{eq:B2+}
\end{eqnarray}
for odd $j\geq 1$.
 
 The integral $I_2$ may be expressed in the form
\begin{equation}
I_2(0,t;m)= h_0+h_1+\frac{1}\lambda (h_2+h_3-h_4-h_5),
\label{eq:EQB15}
\end{equation}
where
\begin{align}
h_0=&\int {\cal D}T \exp \left[(r_+-1)\sum_{k=1}^{m}
t^{\prime}_{k}\right]\int_{\tau_1}^{\tau_2}c({\tau})d\tau \nonumber\\
h_1=&\sideset{}{''}\sum_{j=3}^{2m-1}\int {\cal D}T
\exp\left[(r_+-1)\sum_{k=1}^{m} t^{\prime}_{k}\right]t^{\prime}_{(j+1)/2}
\nonumber\\
h_2=&\sideset{}{''}\sum_{j=3}^{2m-1}\sideset{}{'}\sum_{k=2}^{j-1}\int {\cal
D}T \exp\left[(r_+-1)\sum_{l=1}^{m}
t^{\prime}_{l}+\lambda(\tau_j-\tau_k)\right]\nonumber\\
h_3=&\sideset{}{''}\sum_{j=3}^{2m-1}\sideset{}{''}\sum_{l=1}^{j}\int {\cal D}T
\exp\left[(r_+-1)\sum_{k=1}^{m}
t^{\prime}_{k}+\lambda(\tau_{j+1}-\tau_l)\right]\nonumber\\
h_4=&\sideset{}{''}\sum_{j=3}^{2m-1}\sideset{}{''}\sum_{l=1}^{j}\int {\cal D}T
\exp\left[(r_+-1)\sum_{k=1}^{m} t^{\prime}_{k}+\lambda(\tau_j-\tau_l)\right]
\nonumber\\
h_5=&\sideset{}{''}\sum_{j=3}^{2m-1}\sideset{}{'}\sum_{k=2}^{j-1}\int {\cal
D}T \exp \left[(r_+-1)\sum_{l=1}^{m}
t^{\prime}_{l}+\lambda(\tau_{j+1}-\tau_k)\right].
\label{eq:EQB16}
\end{align}
Obviously, $h_0$ represents $j=1$ in the sum in Eq.~\ref{eq:EQ21}, and hence
includes all possible contributions from $m=1$. The rest of the terms
therefore always have $m\geq 2$. The Laplace transforms for $h_0$ and $h_1$
are straightforward, and one may easily verify that
\begin{align}
{\tilde h}_0(s)= & s^{-(m+1)}(s+1-r_+)^{-m}\times\nonumber \\
 & \quad \left(
 \frac{1}{s+1-r_+}-\frac{1}{s+1-r_{+}+\lambda}\right), \nonumber \\
 {\tilde
 h}_1(s)= & s^{-(m+1)}(s+1-r_+)^{-m}\frac{m-1}{s+1-r_{+}};\quad m\geq 2
\label{eq:EQB17}
\end{align}
so that
\begin{align}
\tilde{h}_0+\tilde{h}_1= & s^{-(m+1)}(s+1-r_+)^{-m}\nonumber \\
 & \quad
 \left( \frac{m}{s+1-r_{+}}-\frac{1}{s+1-r_{+}+\lambda}\right)
\label{eq:EQB18}
\end{align}
for $m\geq 2$. We omit further details of computation of $h_2,h_3,h_4$ and
$h_5$, which involves straightforward, but somewhat tedious algebra. The
results are (for $m\geq 2$)
\begin{align}
{\tilde h}_2&(s)= s^{-(m+1)}(s+1-r_+)^{-m}\times \nonumber \\
 & \qquad
 \left( m+\frac{y^m-1}{1-y}\right)
 \frac{s}{\lambda}\frac{s+1-r_{+}+\lambda}{2s+1-r_{+}+\lambda}\nonumber\\
 {\tilde h}_3&(s)= s^{-(m+1)}(s+1-r_+)^{-m}\times \nonumber \\
 & \left(
 m+\frac{y(y^m-1)}{1-y}-(1-y)\right)
 \left(\frac{s}\lambda+1\right)\frac{s+1-r_{+}}{2s+1-r_{+}+\lambda}\nonumber\\
 {\tilde h}_4& (s)= s^{-(m+1)}(s+1-r_+)^{-m}\times \nonumber \\
 & \left( m+
 \frac{y(y^m-1)}{1-y}-(1-y)\right)
 \left(\frac{s}\lambda+1\right)\frac{s+1-r_{+}+\lambda}{2s+1-r_{+}+\lambda}\nonumber\\
 {\tilde h}_5& (s)= s^{-(m+1)}(s+1-r_+)^{-m}\times \nonumber \\
 & \qquad
 \left( m+\frac{y^m-1}{1-y}\right)
 \frac{s}{\lambda}\frac{s+1-r_{+}}{2s+1-r_{+}+\lambda}.
\label{eq:EQB19}
\end{align}

 After putting together all the terms, we show that
\begin{align}
{\tilde I_2}(s;1)= & s^{-2}(s+1-r_+)^{-1}\nonumber \\
 & \quad \left(
 \frac{1}{s+1-r_+}-\frac{1}{s+1-r_{+}+\lambda}\right)
\label{eq:EQB20}
\end{align}
and for $m\geq 2$,
\begin{eqnarray}
{\tilde I_2}(s;m)=s^{-(m+1)}(s+1-r_+)^{-m}\times~~~~~~~~~~~~~~~\nonumber\\
\left( \frac{m(s+\lambda)}{(s+1-r_+)(2s+1-r_{+}+\lambda)} \right.\nonumber\\
\left. +\frac{y^m-1}{1-y}\frac{s}{(s+1-r_{+}+\lambda)(2s+1-r_{+}+\lambda)}\right).
\label{eq:EQB21}
\end{eqnarray}

%%%%%%%%%%%%%%%%%%%%%%%%%%%%%%%%%%%%%%%%%%%%%%%%%%%%%%%%%%%%%%%%%%%%%%%%%%%%%%%%%

\section{Relation between $\boldsymbol{G_{11}}$ and $\boldsymbol{G_{00}}$ for $\boldsymbol{\alpha \ne 0}$}
\label{sec:appendix_G11}

 In order to proof Eq.~\ref{eq:EQ27} to be valid up to $O(\alpha )$, we
first calculate the functions $f_1(t)$ and $f_2(t)$. As in Eq.~\ref{eq:EQ18},
in Eq.~\ref{eq:EQ32} $c^{\prime}f_1(t)$ and $c^{\prime}f_2(t)$ are the
$O(\alpha)$ terms in $G_{01}(0,c^{\prime};t)$ and $G_{11}(0,c^{\prime};t)$,
respectively, when the channel does not allow entry of ions. The solution is
obtained by solving the corresponding rate equations with the time-dependent
rate $r_{-}(t)=1+\alpha c^{\prime}e^{-\lambda t}$. The results are
\begin{align}
f_1(t)= & -f(t), \nonumber\\
 f_2(t)= &
-\frac{1}{\lambda}\left(\frac{1-\lambda}{1+r_+-\lambda}e^{-(1+r_+)t}
-\frac{1}{1+r_+}e^{-(1+r_++\lambda)t} \right. \nonumber \\
 &
\left. +\frac{\lambda r_+}{(1+r_+)(1+r_+-\lambda)}e^{-\lambda t}\right).
\label{eq:EQ33}
\end{align}

 Instead of Eq.~\ref{eq:EQ28}, we now write a relation between the full
propagators $\Pi$ with $0\le t'\le t$:
\begin{align}
G_{01}(0,0;t) & = \sum_j\int_0^1 dc'\int_0^1 dc
 \Pi_{0j}(0,0;t',c')\Pi_{j1}(t',c';t,c)\nonumber \\
 & = \int_0^1 dc'
 \Pi_{00}(0,0;t',c')G_{01}(t',c';t) \nonumber\\
 & \phantom{=} + \int_0^1 dc'
 \Pi_{01}(0,0;t',c')G_{11}(t',c';t).
\label{eq:G01_propagators}
\end{align}
After substituting Eq.~\ref{eq:EQ32} into Eq.~\ref{eq:G01_propagators}, we
find that
\begin{align}
& G_{01} (0,c_0=0;t)=
 G_{00}(0,c_0=0;t^{\prime})G_{01}(t^{\prime},c^{\prime}=0;t) \nonumber\\
 &
 \; + G_{01}(0,c_0=0;t^{\prime})G_{11}(t^{\prime},c^{\prime}=0;t) \nonumber
 \\
 & \; + \alpha \left( f_{1}(t-t^{\prime})\langle
 c^{\prime}(t')\rangle_{0} + f_2(t-t^{\prime})\langle
 c^{\prime}(t')\rangle_{1}\right) + O(\alpha^2),
\label{eq:EQ34}
\end{align}
where $\langle c^{\prime}(t')\rangle_{0}$ and  $\langle
c^{\prime}(t')\rangle_{1}$ were defined in Eq.~\ref{eq:<c>_01}. Let us now
take the limit of $t^{\prime}\to 0$ as in the case without feedback, and
express the Green's function $G_{11}$ in terms of $G_{00}$.  The result, to
$O(\alpha)$ is
\begin{align}
G_{11} & (0,c_0=0;t) =1-G_{00}(0,0;t) \nonumber\\
 & +
\frac{1}{\left.\partial_t G_{00}(0,0;t)\right|_{t=0}} \big[\partial_t
G_{00}(0,0;t) \nonumber \\
 & -\alpha \left(
f_1(t)\partial_{t^{\prime}}\langle
c^{\prime}(t')\rangle_{0}|_{t^{\prime}=0}+f_2(t)\partial_{t^{\prime}}\langle
c^{\prime}(t')\rangle_{1}|_{t^{\prime}=0}\right) \big].
\label{eq:EQ35}
\end{align}

 It is easy to see now that both the time derivatives of $\langle
c^{\prime}(t')\rangle_{0/1}$ in the previous equation vanish, simply because
they are defined in Eq.~\ref{eq:<c>_01} with an initial condition $c(0)=0$
{\it by construction}. This is easily seen by first solving Eq.~\ref{eq:EQ8}
to find $\langle c(t)\rangle$ to order $\alpha=0$, and the result is
\begin{align}
\langle c(t)\rangle= & \frac{r_{+}}{(1+r_{+})}\Bigg(
 1-\frac{1}{1+r_+-\lambda}\times \nonumber \\
 & \left[(1+r_+)e^{-\lambda
 t}-\lambda e^{-(1+r_+)t}\right]\Bigg) + O(\alpha),
\end{align}
whose time derivative vanishes at $t=0$. But, since $\langle
c(t)\rangle=\langle c(t)\rangle_{0}+\langle c(t)\rangle_{1}$ and since
$\partial_{t}\langle c(t)\rangle_{0/1}|_{t=0}\geq 0$, it follows that
$\partial_{t}\langle c(t)\rangle_{0/1}|_{t=0}=0$ in Eq.~\ref{eq:EQ35} and this
leads to Eq.~\ref{eq:EQ27}.

%%%%%%%%%%%%%%%%%%%%%%%%%%%%%%%%%%%%%%%%%%%%%%%%%%%%%%%%%%%%%%%%%%%%%%%%%%%%%%%%

\section{Calculation of $\boldsymbol{\langle c\rangle_{0}}$ and $\boldsymbol{\langle c\rangle_{1}}$}
\label{sec:appendix_c}

 In this appendix, we compute $\langle c(t)\rangle_{0}$, whose long-time
limit gives $\langle c\rangle_{0}$. Note that, in accordance with
Eq.~\ref{eq:EQ34}, this quantity needs to be computed only to the zeroth order
in $\alpha$. This can be done using the probability functional ${\cal P}_{00}$
from Eq.~\ref{eq:EQ13}. We therefore define
\begin{equation}
\langle c(t)\rangle_{0}=\sum_{m=0}^{\infty}\int {\cal D}T{\cal
P}_{00}[0,t;\{t_i\},\{t^{\prime}_i\};2m]c(t),
\label{eq:EQC2}
\end{equation}
where $c(t)=e^{-\lambda t}\sideset{}{''}\sum_{j=1}^{2m-1}\left(
e^{\lambda\tau_{j+1}}-e^{\lambda\tau_{j}}\right)$ from
Eq.~\ref{eq:EQ8}. Further calculations are done with the aid of Laplace
transforms and using the generalized convolution theorem
(Eq.~\ref{eq:EQB0-}). We omit further details, and simply give the final
result for the Laplace transform $\langle {\tilde c}(s)\rangle_{0}$:
\begin{equation}
\langle {\tilde
c}(s)\rangle_{0}=\frac{r_+\lambda}{s(s+\lambda)(s+1+r_+)(s+1+r_++\lambda)}.
\label{eq:EQC3}
\end{equation}

 The steady state value is obtained after inversion $s\to t$ and taking
the
 limit $t\to\infty$ and is easily found to be
\begin{equation}
\langle c\rangle_{0}=\frac{r_+}{1+r_+}\frac{1}{1+r_++\lambda},
\label{eq:EQC4}
\end{equation}
from which we also find
\begin{equation}
\langle c\rangle_{1}=\langle c\rangle-\langle
c\rangle_{0}=\frac{r_+}{1+r_+}\frac{r_++\lambda}{1+r_++\lambda}.
\label{eq:EQC5}
\end{equation}

\section{List of coefficients in the first order terms}
\label{sec:coefficients}

\subsection{Correlation functions}

\begin{align*}
B_1 = & -\frac{r_+}{(1+r_+)^3(1+r_+-\lambda)^2\lambda(1+r_++\lambda)}\times
 \nonumber\\
 & \left[ -(\lambda-1)r_+^4-(\lambda-1)(2-\lambda)r_+^3 +
 \lambda(\lambda^2-2\lambda-1)r_+^2 \right. \nonumber \\
 &
 \left. +(-\lambda^4+\lambda^3-3\lambda^2+3\lambda-2)r_+ \right.\nonumber\\
 & \left. +(\lambda-1)^2(\lambda+1)(\lambda-1) \right] \nonumber \\
 C_1 = &
 -\frac{2r_+^2\lambda}{(1+r_+)^2(1+r_+-\lambda)^2(1+r_++\lambda)} \nonumber
 \\
 D_1 = & \frac{r_+(r_+^2-1)}{(1+r_+)^3\lambda(1+r_++\lambda)} \nonumber
 \\
 E_1 = & \frac{r_+(\lambda-1)}{(1+r_+)^2(1+r_+-\lambda)} 
\end{align*}

\begin{align*}
B_2 = & -\frac{r_+\lambda}{(1+r_+)^3(1+r_+-\lambda)^3(1+r_++\lambda)^2}
 \times\nonumber\\
 & \left[
 (\lambda-1)r_+^4+(\lambda-1)(2-\lambda)r_+^3-\lambda(\lambda^2-4\lambda+1)r_+^2
 \right. \nonumber \\
 &
 \left. +(\lambda^4-\lambda^3+7\lambda^2-7\lambda+2)r_+ \right.\nonumber\\
 &
 \left. -(\lambda-1)(\lambda^3-\lambda^2-3\lambda+1) \right] \nonumber \\
 C_2 = &
 \frac{r_+\lambda}{(1+r_+)^3(1+r_+-\lambda)^3(1+r_++\lambda)^2(1+r_++2\lambda)}
 \times\nonumber\\
 & \left[ r_+^6 + (1+\lambda)r_+^5 -
 (3\lambda^2-5\lambda+6)r_+^4 \right. \nonumber \\
 &
 \left. +(-\lambda^3+4\lambda^2+6\lambda-14)r_+^3  \right. \nonumber\\
 &
 \left. +(2\lambda^4+\lambda^3+18\lambda^2-2\lambda-11)r_+^2
 \right. \nonumber\\
 &
 \left. +(\lambda^4+3\lambda^3+12\lambda^2-7\lambda-3)r_+
 +\lambda(\lambda^3+\lambda^2+\lambda-3) \right] \nonumber \\
 D_2 = &
 -\frac{r_+\lambda(r_+-1)}{(1+r_+)^3(1+r_++\lambda)(1+r_++2\lambda)}\nonumber
 \\
 E_2 = &
 -\frac{r_+\lambda^2(\lambda-1)}{(1+r_+)^3(1+r_+-\lambda)^3(1+r_++\lambda)^2}\times\nonumber\\
 & \left[ r_+^3 + 3r_+^2 + (3-\lambda^2)r_+ - (\lambda^2-1) \right] \nonumber
 \\
 F_2 = &
 -\frac{r_+^2\lambda^2}{(1+r_+)^2(1+r_+-\lambda)^2(1+r_++\lambda)}
\end{align*}

\subsection{Response functions}

\begin{align*}
B_3 = & -\frac{1}{\lambda(1+r_+)^2(1+r_+-\lambda)^2} \left[ -(\lambda-1)r_+^3
 \right.\nonumber\\
 & \left. +(2\lambda^2-4\lambda+1)r_+^2 +
 (-\lambda^3+2\lambda^2-\lambda-1)r_+  \right.\nonumber\\
 &
 \left. -(\lambda-1)^2 \right] \nonumber \\
 C_3 = &
 -\frac{r_+\lambda}{(1+r_+)^2(1+r_+-\lambda)^2} \nonumber \\
 D_3 = &
 \frac{r_+^2-\lambda-1}{(1+r_+)^2\lambda(1+r_++\lambda)} \nonumber \\
 E_3 =
 & \frac{\lambda-1}{(1+r_+)(1+r_+-\lambda)}
\end{align*}

\begin{align*}
B_4 = & -\frac{1}{(1+r_+)^2(1+r_+-\lambda)^3} \left.\big[ 1 +\lambda
 (2\lambda-3) \right.\nonumber\\
 & \left. +r_+(1+\lambda^2(\lambda-1))-
 r_+^2(1+2\lambda(\lambda-2))+r_+^3(\lambda-1)\right.\big] \nonumber \\
 C_4=
 &
 \frac{\lambda}{(1+r_+)^3(1+r_+-\lambda)^3(1+r_++\lambda)}\times\nonumber\\
 & \left[ \lambda^3(1+r_++r_+^2) \right.\nonumber\\
 &
 \left. -\lambda^2r_+(1+r_+)^2-\lambda(1+r_+)^2(1+r_+(r_+-5))
 \right.\nonumber\\
 & \left. +r_+(r_+-1)(1+r_+)^3\right] \nonumber\\
 D_4 =
 &  -\frac{r_+^2-\lambda-1}{(1+r_+)^3(1+r_++\lambda)} \nonumber \\
 E_4 = &
 -\frac{\lambda(\lambda-1)}{(1+r_+)(1+r_+-\lambda)^2} \nonumber \\
 F_4 = &
 -\frac{r_+\lambda^2}{(1+r_+)^2(1+r_+-\lambda)^2}
\end{align*}

\end{document}